\documentclass{LMCS}

\usepackage{amssymb}
\usepackage{amsmath}
\usepackage{amscd}
\usepackage{latexsym}
\usepackage{epic}
\usepackage{eepic}
\usepackage{psfrag}
\usepackage{psfig}
\usepackage{graphicx}
\usepackage{algorithm}
\usepackage{algorithmic}
\usepackage{luca}
\usepackage{macro}
\usepackage{subfigure}
\usepackage[all]{xy}
\usepackage{hyperref,enumerate}

\theoremstyle{plain}

\def\doi{4 (3:7) 2008}
\lmcsheading%
{\doi}
{1--28}
{}
{}
{Jan.~\phantom{0}4, 2008}
{Sep.~11, 2008}
{}   

\begin{document}

\title[Game Refinement Relations and Metrics]
{Game Refinement Relations and Metrics}

\author[L.~de Alfaro]{Luca de Alfaro\rsuper a}	
\address{{\lsuper a}CE Department, University of California, Santa Cruz}
\email{luca@soe.ucsc.edu}  

\author[R.~Majumdar]{Rupak Majumdar\rsuper b}	
\address{{\lsuper b}Department of CS, University of California, Los Angeles}
\email{rupak@cs.ucla.edu}  

\author[V.~Raman]{Vishwanath Raman\rsuper c}	
\address{{\lsuper c}CS Department, University of California, Santa Cruz}
\email{vishwa@soe.ucsc.edu}  

\author[M.~Stoelinga]{Mari\"elle Stoelinga\rsuper d}	
\address{{\lsuper d}Department of CS, University of Twente, The Netherlands}
\email{marielle@cs.utwente.nl}  



\keywords{game semantics, minimax theorem, metrics, $\omega$-regular
properties, quantitative $\mu$-calculus, probabilistic choice, 
equivalence of states, refinement of states}


\subjclass{F.4.1, F.1.1} 

\titlecomment{{\lsuper *}A version of this paper titled ``Game Relations and
Metrics" appeared in the $22^{nd}$ Annual IEEE Symposium on Logic in
Computer Science, July 2007}



\begin{abstract}
\noindent We consider two-player games played over finite state spaces for an
infinite number of rounds. 
At each state, the players simultaneously choose moves; the moves
determine a successor state.
It is often advantageous for players to choose
probability distributions over moves, rather than single moves.
Given a goal (e.g., ``reach a target state''), the question of winning
is thus a probabilistic one: ``what is the maximal probability of
winning from a given state?''. 

On these game structures, two fundamental notions are those of 
{\em equivalences\/} and {\em metrics}. 
Given a set of winning conditions, two states are {\em equivalent\/}
if the players can win the same games with the same probability from
both states.  
{\em Metrics\/} provide a bound on the difference in the probabilities
of winning across states, capturing a quantitative notion of
state ``similarity''.

We introduce equivalences and metrics for two-player game structures,
and we show that they characterize the difference in probability of
winning games whose goals are expressed in the quantitative $\mu$-calculus. 
The quantitative $\mu$-calculus can express a large set of goals,
including reachability, safety, and $\omega$-regular properties. 
Thus, we claim that our relations and metrics provide the canonical
extensions to games, of the classical notion of bisimulation for
transition systems. 
We develop our results both for equivalences and metrics, which
generalize bisimulation, and for asymmetrical versions, which
generalize simulation. 
\end{abstract}

\maketitle

\vfill
\section{Introduction}

We consider two-player games played for an infinite number of rounds
over finite state spaces. 
At each round, the players simultaneously and independently select
moves; the moves then determine a probability distribution over
successor states. 
These games, known variously as {\em stochastic games\/}
\cite{Shapley53} or {\em concurrent\/} games
\cite{luca-focs98,ATL02,dAM04}, generalize many common structures in
computer science, from transition systems, to Markov chains
\cite{Kemeny} and Markov decision processes \cite{Derman}. 
The games are {\em turn-based\/} if, at each state, at most one
of the players has a choice of moves, and {\em deterministic\/} if the
successor state is uniquely determined by the current state, and by
the moves chosen by the players. 

It is well-known that in such games with simultaneous moves it is
often advantageous for the players to randomize their moves, so that
at each round, they play not a single ``pure'' move, but rather, a
probability distribution over the available moves. 
These probability distributions over moves, called {\em mixed moves\/}
\cite{OsborneRubinstein}, lead to various notions of
equilibria \cite{vonNeumannMorgenstern44,OsborneRubinstein}, such as
the equilibrium result expressed by the minimax theorem
\cite{vonNeumannMorgenstern44}. 
Intuitively, the benefit of playing mixed, rather than pure, moves
lies in preventing the adversary from tailoring a response to the
individual move played. 
Even for simple reachability games, the use of mixed moves may allow
players to win, with probability~1, games that they would lose (i.e.,
win with probability~0) if restricted to playing pure moves 
\cite{luca-focs98}. 
With mixed moves, the question of winning a game with respect to a 
goal is thus a probabilistic one: what is the maximal probability a player
can be guaranteed of winning, regardless of how the other player
plays? 
This probability is known, in brief, as the {\em winning probability.}

In structures ranging from transition systems to Markov decision
processes and games, a fundamental question is the one of 
{\em equivalence\/} of states.
Given a suitably large class $\Phi$ of properties, containing all
properties of interest to the modeler, two states are equivalent if the same
properties hold in both states. 
For a property $\varphi$, denote the value of $\varphi$ at $s$ by $\varphi(s)$: 
in the case of games, this might represent the maximal probability of a
player winning with respect to a goal expressed by $\varphi$. 
Two states $s$ and $t$ are equivalent if $\varphi(s) = \varphi(t)$ for
all $\varphi \in \Phi$.
For 
(finite-branching) 
transition systems, and for the class of properties $\Phi$
expressible in the $\mu$-calculus \cite{Kozen83mu}, state
equivalence is captured by bisimulation \cite{Milner90}; 
for Markov decision processes, it is captured by probabilistic
bisimulation \cite{SL94}. 
For quantitative properties, a notion related to equivalence is that
of a {\em metric:\/} a metric provides a tight bound for how much the
value of a property can differ at states of the system, and provides
thus a quantitative notion of similarity between states. 
Given a set $\Phi$ of properties, the metric distance of two
states $s$ and $t$ can be defined as $\sup_{\varphi \in \Phi} |\varphi(s)
- \varphi(t)|$. 
Metrics for Markov decision processes have been studied in 
\cite{DGJP99,vanBreugelCONCUR01,vanBreugel-icalp2001,DGJP02,RadhaLICS02}.
Obviously, the metrics and relations are connected, in the sense that
the relations are the {\em kernels\/} of the metrics (the pairs of
states having metric distance~0). 
The metrics and relations are at the heart of many verification
techniques, from approximate reasoning (one can substitute states that
are close in the metric) to system reductions (one can collapse
equivalent states) to compositional reasoning and refinement
(providing a notion of substitutivity of equivalents). 

We introduce metrics and equivalence relations for
concurrent games, with respect to the class of properties $\Phi$
expressible in the quantitative $\mu$-calculus
\cite{dAM04,IverMorgan}.  
We claim that these metrics and relations represent the canonical
extension of bisimulation to games. 
We also introduce asymmetrical versions of these metrics and
equivalences, which constitute the canonical extension of simulation. 

An equivalence relation for deterministic games that are either
turn-based, or where the players are constrained to playing pure
moves, has been introduced in \cite{CONCUR98AHKV} and called
{\em alternating bisimulation.} 
Relations and metrics for the general case of concurrent games have so
far proved elusive, with some previous attempts at their definition by 
a subset of the authors following a subtly flawed approach
\cite{luca-icalp-disc-03,rupak-thesis}. 
The cause of the difficulty  goes to the heart of
the definition of bisimulation. 
In the definition of bisimulation for transition systems, for every
pair $s$, $t$ of bisimilar states, we require that if $s$ can go to a
state $s'$, then $t$ should be able to go to $t'$, such that $s'$ and
$t'$ are again bisimilar (we also ask that $s$, $t$ have an equivalent
predicate valuation). 
This definition has been extended to Markov decision processes by
requiring that for every mixed move from $s$, there is a mixed move
from $t$, such that the moves induce probability distributions over
successor states that are equivalent modulo the underlying
bisimulation \cite{SL94,segalaT}. 
Unfortunately, the generalization of this appealing definition to
games fails. 
It turns out, as we prove in this paper, that requiring players to be
able to replicate probability distributions over successors (modulo
the underlying equivalence) leads to an equivalence that is too fine,
and that may fail to relate states at which the same quantitative
$\mu$-calculus formulas hold. 
We show that phrasing the definition in terms of distributions over
successor states is the wrong approach for games; rather, the
definition should be phrased in terms of expectations of certain
metric-bounded quantities. 

Our starting point is a closer
look at the definition of metrics for Markov decision processes. 
We observe that we can manipulate the definition of metrics given in
\cite{vanBreugel-icalp2001}, obtaining an alternative form, which we
call the {\em a priori\/} form, in contrast with the original form of
\cite{vanBreugel-icalp2001}, which we call the {\em a posteriori\/}
form. 
Informally, the a posteriori form is the traditional definition,
phrased in terms of similarity of probability distributions; 
the a priori form is instead phrased in terms of expectations. 
We show that, while on Markov decision processes these two forms
coincide, this is not the case for games; moreover, we show that it is
the a priori form that provides the canonical metrics for
games. 

We prove that the a priori metric distance between two
states $s$ and $t$ of a concurrent game is equal to 
$\sup_{\varphi \in \Phi} |\varphi(s) - \varphi(t)|$, where $\Phi$ is the set of
properties expressible via the quantitative $\mu$-calculus. 
This result can be summarized by saying that the quantitative $\mu$-calculus
provides a {\em logical characterization\/} for the a priori metrics,
similar to the way the ordinary $\mu$-calculus provides a
logical characterization of bisimulation. 
Furthermore, we prove that a priori metrics --- and their kernels, the
a priori relations --- satisfy a {\em reciprocity\/} property, stating
that properties expressed in terms of player~1 and player~2 winning
conditions have the same distinguishing power. 
This property is intimately connected to the fact that concurrent
games, played with mixed moves, are {\em determined\/} for
$\omega$-regular goals \cite{Martin98,dAM04}: the probability that
player~1 achieves a goal $\psi$ is one minus the probability that
player~2 achieves the goal $\no \psi$. 
Reciprocity ensures that there is one, canonical, notion of game
equivalence.  
This is in contrast to the case of alternating bisimulation of
\cite{CONCUR98AHKV}, in which there are distinct player~1 and player~2
versions, as a consequence of the fact that concurrent games, when
played with pure moves, are not determined. 
The logical characterization and reciprocity result justify our claim
that a priori metrics and relations are the canonical notion of
metrics, and equivalence, for concurrent games. 
Neither the logical characterization nor the reciprocity result hold
for the a posteriori metrics and relations. 

While this introduction focused mostly on metrics and equivalence
relations, we also develop results for the asymmetrical
versions of these notions, related to simulation.

\section{Games and Goals}

We will develop metrics for game structures over
a set $S$ of states. 
We start with some preliminary definitions.
For a finite set $A$, let $\distr(A)=\{p:A\mapsto[0,1]\mid \sum_{a\in
A}p(a)=1\}$ denote the set of probability distributions over $A$. 
We say that $p \in \distr(A)$ is {\em deterministic\/} if there is 
$a \in A$ such that $p(a) = 1$. 

For a set $S$, a {\em valuation over $S$\/} 
is a function $\valu: S \mapsto [0,1]$ associating with every element $s
\in S$ a value $0 \leq \valu(s) \leq 1$; we let $\valus$ be the set of
all valuations. 
For $c \in [0,1]$, we denote by $\imeanbb{c}$ the constant valuation
such that $\imeanbb{c}(s) = c$ at all $s \in S$. 
We order valuations pointwise: for $\valu, \valub \in \valus$, we
write $\valu \leq \valub$ iff $\valu(s) \leq \valub(s)$ at all $s \in S$;
we remark that $\valus$, under $\leq$, forms a complete lattice. 

Given $a,b \in \reals$, we write $a \imax b = \max\set{a,b}$, and 
$a \imin b = \min\set{a,b}$; we also let 
$a \oplus b = \min \set{1, \max \set{0, a + b}}$ and
$a \ominus b = \max \set{0, \min \set{1, a - b}}$. 
We extend $\imin, \imax, +, -, \oplus, \ominus$ to valuations by interpreting
them in pointwise fashion. 

A {\em directed metric\/} is a function $d: S^2 \mapsto \reals_{\geq 0}$ 
which satisfies $d(s,s) = 0$ and $d(s,t) \leq d(s,u) + d(u,t)$ for all 
$s,t,u \in S$. 
We denote by $\metrsp \subseteq S^2 \mapsto \reals$ the space of all metrics;
this space, ordered pointwise, forms a lattice which we indicate with
$(\metrsp, \leq)$. 
Given a metric $d \in \metrsp$, we denote by $\breve{d}$ its 
{\em opposite\/} version, defined by $\breve{d}(s,t) = d (t,s)$ for
all $s, t \in S$; we say that $d$ is symmetrical if $d = \breve{d}$.

\subsection{Game Structures}

We assume a fixed, finite set $\vars$ of {\em observation variables}.
A  (two-player, concurrent) {\em game structure\/} 
$\game=\tuple{S,\int{\cdot},\moves,\mov_1,\mov_2,\trans}$
consists of the  following components \cite{ATL02,luca-focs98}:
\begin{enumerate}[$\bullet$]

\item A finite set $S$ of states.

\item A variable interpretation $\int{\cdot}: \vars \times S
  \mapsto [0,1]$, which associates with each variable $\varx \in
  \vars$ a valuation $\int{v}$. 

\item A finite set $\moves$ of moves.

\item Two move assignments 
  $\mov_1,\mov_2$: $S\mapsto 2^\moves\setm\emptyset$.  
  For $i\in\{1,2\}$, the assignment $\mov_i$ associates with each
  state $s \in S$ the nonempty set $\mov_i(s)\subseteq\moves$ of moves
  available to player~$i$ at state~$s$.

\item A probabilistic transition function $\trans$: 
  $S\times\moves\times \moves\mapsto\distr(S)$, that gives the probability 
  $\trans(s,a_1,a_2)(t)$ of a transition from $s$ to $t$ when player~1
  plays move $a_1$ and player~2 plays move~$a_2$. 
\end{enumerate}
At every state $s\in S$, player~1 chooses a move $a_1\in\mov_1(s)$, 
and simultaneously and independently player~2 chooses a move 
$a_2\in\mov_2(s)$.  
The game then proceeds to the successor state $t\in S$ with probability
$\trans(s,a_1,a_2)(t)$.
We denote by $\dest(s,a_1,a_2) = \set{t \in S \mid \trans(s,a_1,a_2)(t) > 0}$
the set of {\em destination states\/} when actions $a_1, a_2$ are
chosen at $s$. 
The variables in $\vars$ naturally induce an equivalence on
states: for states $s,t$, define $s\loceq t$ if for all $\varx\in\vars$ we have
$\int{\varx}(s) = \int{\varx}(t)$.
In the following, unless otherwise noted, the definitions refer to a game
structure with components 
$\game=\tuple{S,\int{\cdot},\moves,\mov_1,\mov_2,\trans}$. 
For player $\ii\in\set{1,2}$, we write $\jj = 3 - \ii$ for the opponent.
We also consider the following subclasses of game structures. 
\begin{enumerate}[$\bullet$]
\item{{\em Turn-based game structures.} 
A game structure $\game$ is {\em turn-based\/} if we can
write $S$ as the disjoint union of two sets: the set $S_1$ of {\em
player~1 states,} and the set $S_2$ of {\em player~2 states,} such
that $s \in S_1$ implies $|\mov_2(s)| = 1$, and $s \in S_2$ implies
$|\mov_1(s)| = 1$, and 
further,
there is a special variable
$\turn \in \vars$, such that $\int{\turn}{(s)} = 1$ iff $s \in S_1$, and 
$\int{\turn}{(s)} = 0$ iff $s \in S_2$: thus, the variable $\turn$
indicates whose turn it is to play at a state. 
}
\item{{\em Markov decision processes.}
A game structure $\game$ is a {\em Markov decision process\/}  (MDP)
\cite{Derman}
if only one of the two players has a choice of moves. 
For $\ii \in \set{1,2}$, we say that a structure is an $\ii$-MDP
if $\forall s \in S$, $\vert \mov_{\jj}(s) \vert = 1$. 
For MDPs, we omit the (single) move of the player
without a choice of moves, and 
write $\delta(s,a)$ for the transition function. 
}
%
\item{{\em Deterministic game structures.} 
A game structure $\game$ is {\em deterministic\/} if, for
all $s \in S$, $a_1 \in \moves$, and $a_2 \in \moves$, there
exists a $t \in S$ such that $\trans(s, a_1, a_2)(t) = 1$; we denote
such $t$ by $\ddest(s,a_1,a_2)$.
We sometimes call {\em probabilistic\/} a general game structure, to
emphasize the fact that it is not necessarily deterministic. 
}
\end{enumerate}
Note that MDPs can be seen as turn-based games by
setting $\int{\turn} = {\bf 1}$ for $1$-MDPs and $\int{\turn} = {\bf 0}$ 
for $2$-MDPs.

\smallskip\noindent{\bf Pure and mixed moves.}
A {\em mixed move\/} is a probability distribution over the moves
available to a player at a state. 
We denote by $\dis_i(s) = \distr(\mov_i(s))$ the set of mixed moves 
available to player~$i \in \set{1,2}$ at $s \in S$. 
The moves in $\moves$ are called {\em pure moves,} in
contrast to mixed moves.
We extend the transition function to mixed moves.
For $s \in S$ and $x_1 \in \dis_1(s)$, $x_2 \in \dis_2(s)$, we write
$\trans(s,x_1,x_2)$ for the next-state probability distribution
induced by the mixed moves $x_1$ and $x_2$, defined for all $t \in S$
by
\begin{align*}
  \trans(s,x_1,x_2)(t) = 
  \sum_{a_1 \in \mov_1(s)} \: 
  \sum_{a_2 \in \mov_2(s)} \:
  \trans(s,a_1,a_2)(t) \: x_1(a_1) \: x_2(a_2) \eqpun . 
\end{align*}
In the following, we sometimes restrict the moves of the players to pure moves.
We identify a pure move $a\in\mov_i(s)$ available to player $i\in\set{1,2}$
at a state $s$ with a deterministic distribution that plays $a$ with probability $1$.

\smallskip\noindent{\bf The deterministic setting.}
The {\em deterministic setting\/} is obtained by considering
deterministic game structures, with players restricted to playing pure
moves. 

\paragraph{Predecessor operators.}
Given a valuation $\valu \in \valus$, a state $s \in S$, 
and two mixed moves $x_1 \in \dis_1(s)$ and $x_2 \in \dis_2(s)$, we define
the expectation of $\valu$ from $s$ under $x_1, x_2$:
\[
  \E^{x_1,x_2}_s (\valu) = 
  \sum_{t \in S} \, \trans(s,x_1,x_2)(t) \, \valu(t) \eqpun .
\]
For a game structure $\game$, for $i \in \set{1,2}$ we define the 
{\em valuation transformer\/} $\pre_i: \valus \mapsto \valus$ 
by, for all $\valu \in \valus$ and $s \in S$, 
\[
  \pre_i(\valu)(s) =
  \Sup_{x_\ii \in \dis_\ii(s)} \; \Inf_{x_\jj \in \dis_\jj(s)} 
  \E^{x_1,x_2}_s (\valu) \eqpun . 
\]
Intuitively, $\pre_i(\valu)(s)$ is the maximal expectation player $i$
can achieve of $\valu$ after one step from $s$: this is the classical
``one-day'' or ``next-stage'' operator of the theory of repeated games
\cite{FilarVrieze97}. 
We also define a {\em deterministic\/} version of this operator, in
which players are forced to play pure moves: 
\[
  \dpre_i(\valu)(s) = \textstyle 
  \Max_{x_\ii \in \ddis_\ii(s)} \; \Min_{x_\jj \in \ddis_\jj(s)} 
  \E^{x_1,x_2}_s (\valu) \eqpun . 
\]

\subsection{Quantitative $\mu$-calculus}

We consider the set of properties expressed by the {\em
quantitative $\mu$-calculus\/} ($\qmu$). 
As discussed in 
\cite{Kozen83,dAM04,IverMorgan}, a large set of properties
can be encoded 
in $\qmu$, spanning from basic properties such as maximal reachability
and safety probability, to the maximal probability of satisfying a
general $\omega$-regular specification. 

\smallskip
\noindent{\bf Syntax.}
The syntax of quantitative $\mu$-calculus is defined with respect to
the set of observation variables $\vars$ as well as 
a set $\mvars$ of {\em calculus variables,} which are
distinct from the observation variables in $\vars$. 
The syntax is given as follows: 
\begin{align*}
  \varphi\ ::=\ &
    c \mid 
    \varx \mid 
    \mvar \mid
    \neg \varphi \mid 
    \varphi\vee\varphi \mid 
    \varphi \wedge\varphi \mid  
    \varphi \oplus c \mid
    \varphi \ominus c \mid
    \qpre_1(\varphi) \mid 
    \qpre_2(\varphi) \mid
    \mu \mvar.\, \varphi \mid
    \nu \mvar.\, \varphi
\end{align*}
for constants $c \in [0,1]$, observation variables $\varx \in \vars$, 
and calculus variables $\mvar \in \mvars$. 
In the formulas $\mu \mvar.\, \varphi$ and $\nu \mvar.\, \varphi$, we
furthermore require that all occurrences of the bound variable $\mvar$ in $\varphi$ occur in
the scope of an even number of occurrences of the complement operator
$\neg$. 
A formula $\varphi$ is {\em closed\/} if every calculus variable $\mvar$
in $\varphi$ occurs in the scope of a quantifier $\mu \mvar$ or $\nu
\mvar$.
From now on, with abuse of notation, we denote by $\qmu$ the set of closed
formulas of $\qmu$.  
A formula is a {\em player~$i$ formula,} for $i \in \set{1,2}$, 
if $\varphi$ does not contain the $\qpre_{\jj}$ operator; we denote with
$\qmu_i$ the syntactic subset of $\qmu$ consisting only of closed player~$i$
formulas. 
A formula is in {\em positive form\/} if the negation appears only in
front of observation variables, i.e., in the context $\neg \varx$; we 
denote with $\qmu^{+}$ and $\qmu_i^{+}$ the subsets of $\qmu$ and $\qmu_i$
consisting only of positive formulas.  

We remark that the fixpoint operators $\mu$ and $\nu$ will not be
needed to achieve our results on the logical characterization of game
relations. 
They have been included in the calculus because they allow the
expression of many interesting properties, such as safety,
reachability, and in general, $\omega$-regular properties. 
The operators $\oplus$ and $\ominus$, on the other hand, are 
necessary for our results. 

\smallskip\noindent
{\bf Semantics.}
A variable valuation  $\xenv$: $\mvars\mapsto\valus$ is a function that maps 
every variable $\mvar\in\mvars$ to a valuation in~$\valus$.
We write $\xenv[\mvar\mapsto\valu]$ for the valuation that agrees
with $\xenv$ on all variables, except that $\mvar$ is mapped to~$\valu$. 
Given a game structure $\game$ and a variable valuation $\xenv$, every formula $\varphi$ of the
quantitative $\mu$-calculus defines a valuation 
$\sem{\varphi}^\game_{\xenv}\in\valus$ (the superscript $\game$ is
omitted if the game structure is clear from the context):
\begin{align*}
  & \sem{c}_{\xenv} = \imeanbb{c}  \\
  & \sem{\varx}_{\xenv} = \int{\varx} \\
  & \sem{\mvar}_{\xenv} = \xenv(\mvar) \\
  & \sem{\neg \varphi}_\xenv = \imeanbb{1} - \sem{\varphi}_\xenv \\
  & \textstyle \sem{\varphi {\oplus \brace \ominus} c}_\xenv = \textstyle \sem{\varphi}_\xenv {\oplus \brace \ominus} \imeanbb{c} \\
  & \textstyle \sem{\varphi_1\,{\vee\brace\wedge}\,\varphi_2}_{\xenv} =
    \textstyle \sem{\varphi_1}_{\xenv} \,{\imax\brace\imin}\, 
    \sem{\varphi_2}_{\xenv} \\
  & \sem{\qpre_i(\varphi)}_\xenv = \pre_i(\sem{\varphi}_\xenv) \\
  & \textstyle \sem{{\mu\brace\nu}\mvar.\, \varphi}_{\xenv} =
	\textstyle {\inf\brace\sup}\set{\valu \in \valus \mid 
    	\valu=\sem{\varphi}_{\xenv[\mvar\mapsto \valu]}}
\end{align*}
where $i \in \set{1,2}$. 
The existence of the fixpoints is guaranteed by the monotonicity 
and continuity of all operators and can be computed by Picard iteration \cite{dAM04}.
If $\varphi$ is closed, $\sem{\varphi}_\xenv$ is independent of $\xenv$, and
we write simply $\sem{\varphi}$. 
%

We also define a {\em deterministic\/} semantics $\dsem{\cdot}$ for
$\qmu$, in which players can select only pure moves in the
operators $\qpre_1$, $\qpre_2$. 
$\dsem{\cdot}$ is defined as $\sem{\cdot}$, except for the clause 
\[
   \dsem{\qpre_i(\varphi)}_\xenv = \dpre_i(\dsem{\varphi}_\xenv) \eqpun . 
\]

\begin{examp}{}
Given a set $T \subs S$, the {\em characteristic valuation\/}
$\imeanbb{T}$ of $T$ is defined by $\imeanbb{T}(s) = 1$ if $s \in T$, 
and $\imeanbb{T}(s) = 0$ otherwise. 
With this notation, the maximal probability with which
player~$i\in\set{1,2}$ can ensure eventually reaching $T \subs S$ is given by
$\sem{\mu \mvar . (\imeanbb{T} \oder \qpre_i(\mvar))}$, 
and the maximal probability with which player~$i$ can guarantee
staying in $T$ forever is given by
$\sem{\nu \mvar . (\imeanbb{T} \und \qpre_i(\mvar))}$ (see, e.g., \cite{dAM04}). 
The first property is called a {\em reachability\/} property, 
the second a {\em safety\/} property. 
\end{examp}

\section{Metrics}

We are interested in developing a {\em metric\/} on states of a game
structure that captures an approximate notion of equivalence: states
close in the metric should yield similar values to the players for any winning
objective.
Specifically, we are interested in defining a bisimulation metric $[\altbis_g] \in \metrsp$
such that 
for any game structure $\game$ and states $s,t$ of $\game$,
the following continuity property holds:
\begin{equation} \label{eq-desired-chacact}
  [\altbis_g] (s,t) = \sup_{\varphi \in \qmu} 
  |\sem{\varphi}(s) - \sem{\varphi}(t)| \eqpun .
\end{equation}
In particular, the kernel of the metric, that is, states at distance $0$,
are equivalent: each player can get exactly the same value from either state
for any objective.
Notice that in defining the metric independent of a player,
we are expecting our metrics to be {\em reciprocal}, that is, invariant
under a change of player. 
Reciprocity is expected to hold since the underlying games we consider 
are determined ---for any game, the value obtained by player 2 is one minus the value obtained
by player 1--- and yields canonical metrics on games.

Thus, our metrics will generalize equivalence and refinement relations that
have been studied on MDPs and in the deterministic setting.
To underline the connection between classical equivalences
and the metrics we develop, we write $[s \altbis_g t]$ 
for $[\altbis_g](s,t)$, so that the desired property
of the bisimulation metric can be stated as 
\[
  [s \altbis_g t] = \sup_{\varphi \in \qmu} 
  |\sem{\varphi}(s) - \sem{\varphi}(t)| \eqpun .
\]
Metrics of this type have already been developed for Markov decision
processes (MDPs) \cite{vanBreugelCONCUR01,DGJP02}. 
Our construction of metrics for games starts from an analysis of
these constructions.

\subsection{Metrics for MDPs}

We consider the case of 1-MDPs; the case for 2-MDPs is symmetrical.
Throughout this subsection, we fix a 1-MDP 
$\tuple{S,\int{\cdot},\moves,\mov_1,\mov_2,\trans}$. 
Before we present the metric correspondent of probabilistic simulation,
we first rephrase classical probabilistic (bi)simulation on MDPs
\cite{CompProb92,JouSmol90,SL94,SL95} as a  
fixpoint of a {\em relation transformer.}
As a first step, we lift relations between states to relations between
distributions.
Given a relation $R \subs S \times S$ and two 
distributions $p, q \in \distr(S)$, we let 
$p \sqsubseteq_{R} q$ if there is a function 
$\Delta : S\times S \rightarrow [0,1]$
such that:
\begin{enumerate}[$\bullet$]
\item $\Delta(s,s') > 0$ implies $(s,s')\in R$;
\item $p(s) = \sum_{s'\in S} \Delta(s,s')$ for any $s\in S$;
\item $q(s') = \sum_{s\in S} \Delta(s,s')$ for any $s'\in S$.
\end{enumerate}
To rephrase probabilistic simulation, we define the relation
transformer $F: 2^{S \times S} \mapsto 2^{S \times S}$ as follows. 
For all relations $R \subs S \times S$ and $s,t \in S$, 
we let $(s,t) \in F(R)$ iff
\begin{equation} \label{eq-mdp-rel-fix}
  s \loceq t  \; \und \; 
   \forall x_1 {\in} \dis_1(s) \qdot \exists y_1 {\in} \dis_1(t) \qdot 
   \trans(s,x_1) \sqsubseteq_{R} \trans(t,y_1),
\end{equation}
for all states $s, t \in S$. 
Probabilistic simulation is the greatest fixpoint of
(\ref{eq-mdp-rel-fix}); probabilistic bisimulation is the greatest
symmetrical fixpoint of (\ref{eq-mdp-rel-fix}). 

To obtain a metric equivalent of probabilistic simulation, we lift the
above fixpoint from relations (subsets of $S^2$) to metrics
(maps $S^2 \mapsto \reals$). 
First, we define $\metr{\loceq}\in\metrsp$ for all $s, t \in S$ by 
$
  \metr{s \loceq t} 
  = \max_{\varx \in \vars} |\int{\varx}(s) - \int{\varx}(t)|
$.
Second, we lift (\ref{eq-mdp-rel-fix}) to metrics, defining a 
metric transformer $\Hmdppost: \metrsp \mapsto \metrsp$. 
For all $d \in \metrsp$, let 
$D(\trans(s,x_1),\trans(t,y_1))(d)$ be the {\em distribution distance} between 
$\trans(s,x_1)$ and $\trans(t,y_1)$ with respect to the metric $d$. 
We will show later how to define such a distribution distance. 
For $s, t \in S$, we let 
\begin{align}
  \Hmdppost(d)(s,t) = \metr{s \loceq t} \imax 
  \Sup_{x_1 \in \dis_1(s)} \,
  \Inf_{y_1 \in \dis_1(t)} D(\trans(s,x_1),\trans(t,y_1))(d) \eqpun .
  \label{eq-mdp-met-fix}
\end{align}
In this definition, the $\forall$ and $\exists$ of
(\ref{eq-mdp-rel-fix}) have been replaced by $\Sup$ and $\Inf$,
respectively.
Since equivalent states should have distance~0, the simulation metric
in MDPs is defined as the {\em least\/} (rather than greatest)
fixpoint of (\ref{eq-mdp-met-fix}) \cite{vanBreugelCONCUR01,DGJP02}. 
Similarly, the bisimulation metric is defined as the least symmetrical
fixpoint of (\ref{eq-mdp-met-fix}).

For a distance $d \in \metrsp$ and two distributions $p, q \in
\distr(S)$, the {\em distribution distance\/} $D(p,q)(d)$ 
is a measure of how much ``work'' we have to
do to make $p$ look like $q$, given that moving a unit of probability
mass from $s \in S$ to $t \in S$ has cost $d(s,t)$. 
More precisely, $D(p,q)(d)$ is defined via the {\em trans-shipping problem,}
as the minimum cost of shipping the distribution $p$ into $q$, with
edge costs $d$. 
Thus, $D(p,q)(d)$ is the solution of the following linear programming (LP)
problem over the set of variables $\set{\lambda_{s,t}}_{s, t \in S}$:
\begin{gather*}
\mathrm{Minimize} \quad  \sum_{s, t \in S}d(s,t) \lambda_{s,t} \\[1ex]
\mathrm{subject\ to\ } 
\sum_{t \in S} \lambda_{s,t} = p(s), 
\quad 
\sum_{s \in S} \lambda_{s,t} = q(t), 
\quad 
\lambda_{s,t} \ge 0 \eqpun .
\end{gather*}
Equivalently, we can define $D(p,q)(d)$ via the dual of the above LP
problem \cite{vanBreugelCONCUR01}.
Given a metric $d \in \metrsp$, let $C(d) \subs \valus$ be the subset
of valuations $k \in \valus$ such that $k(s) - k(t) \leq d(s, t)$ for
all $s, t \in S$. 
Then the dual formulation is:
\begin{gather}\label{eq-lp}
\mathrm{Maximize} \quad    \sum_{s \in S} p(s)\, k(s) - 
                     \sum_{s \in S} q(s) k(s) \\[1ex] 
\mathrm{subject\ to\ }  k \in C(d) \eqpun . \nonumber 
\end{gather}
The constraint $C(d)$ on the valuation $k$, states that the value of $k$
across states cannot differ by more than $d$. 
This means, intuitively, that $k$ behaves like the valuation of a
$\qmu$ formula: as we will see, the logical characterization implies
that $d$ is a bound for the difference in valuation of $\qmu$ formulas
across states. 
Indeed, the logical characterization of the metrics is proved by
constructing formulas whose valuation approximate that of the optimal
$k$. Plugging (\ref{eq-lp}) into (\ref{eq-mdp-met-fix}), we obtain: 
\begin{align} 
  \Hmdppost(d)(s,t)  = \metr{s \loceq t} \; \imax
  \Sup_{x_1 \in \dis_1(s)} \, \Inf_{y_1 \in \dis_1(t)} \,
  \Sup_{k \in C(d)}
  \bigl(
  \E_{s}^{x_{1}}(k) - \E_{t}^{y_{1}}(k)
  \bigr) \eqpun .
  \label{eq-post-metric}
\end{align}
We can interpret this definition as follows.
State $t$ is trying to simulate state $s$ (this is a definition of a
simulation metric). 
First, state $s$ chooses a mixed move $x_1$, attempting to make
simulation as hard as possible; then, state $t$ chooses a mixed move 
$y_1$, trying to match the effect of $x_1$. 
Once $x_1$ and $y_1$ have been chosen, the resulting 
distance between $s$ and $t$ is equal to the maximal difference in
expectation, for moves $x_1$ and $y_1$, of a valuation $k \in C(d)$.
We call the metric transformer $\Hmdppost$ the {\em a posteriori\/}
metric transformer: the valuation $k$ in (\ref{eq-post-metric})
is chosen {\em after\/} the moves $x_1$ and $y_1$ are chosen. 
We can define an {\em a priori\/} metric transformer, where $k$ is
chosen before $x_1$ and $y_1$: 
\begin{align} 
  \Hmdpprio(d)(s,t) = \metr{s \loceq t} \; \imax
  \Sup_{k \in C(d)} \,
  \Sup_{x_1 \in \dis_1(s)} \, \Inf_{y_1 \in \dis_1(t)}
  \bigl(
  \E_{s}^{x_{1}}(k) - \E_{t}^{y_{1}}(k)
  \bigr) \eqpun .
  \label{eq-prio-metric}
\end{align}
Intuitively, in the a priori transformer, first a valuation $k \in C(d)$
is chosen. 
Then, state $t$ must simulate state $s$ with respect to the
expectation of $k$. 
State $s$ chooses a move $x_1$, trying to maximize the difference in
expectations, and state $t$ chooses a move $y_1$, trying to minimize
it. 
The distance between $s$ and $t$ is then equal to the difference in
the resulting expectations of $k$. 

Theorem~\ref{thm-mpdpriovspost} below states that for MDPs, a priori
and a posteriori simulation metrics coincide.
In the next section, we will see that this is not the case for games. 

\begin{thm}{}\label{thm-mpdpriovspost}
For all MDPs, 
$\Hmdppost = \Hmdpprio$. 
\end{thm}

\proof
Consider two states $s, t \in S$, and a metric $d \in \metrsp$.
We have to prove that 

\begin{gather}
\sup_{k} \, \sup_{x_{1}} \, \inf_{y_{1}} [\E_s^{x_1}(k) - \E_t^{y_1}(k)]
= \sup_{x_{1}} \, \inf_{y_{1}} \, \sup_{k} [\E_s^{x_1}(k)-\E_t^{y_1}(k)]
\eqpun . 
\end{gather}

In the left-hand side, we can exchange the two 
outer $\sup$s. Then, noticing that the difference in expectation is bi-linear 
in $k$ and $y_1$ for a fixed $x_1$, that $y_1$ is a probability 
distribution, and that $k$ is chosen from a compact
convex subset, we apply the generalized minimax theorem \cite{Sio58}
to exchange 
$\sup_k \inf_{y_{1}}$ into $\inf_{y_{1}} \sup_k$, thus obtaining the
right-hand side.\qed


The metrics defined above are logically characterized by $\qmu$. 
Precisely, let $[\eqmdp] \in \metrsp$ be the least symmetrical fixpoint of
$\Hmdpprio = \Hmdppost$. 
Then, Lemma~$5.24$ and Corollary~$5.25$ of \cite{DGJP02}, 
(originally stated for $\Hmdppost$)
state that for all states $s, t$ of a 1-MDP, we have 
\[
  [s \eqmdp t] = \sup_{\varphi \in \qmu} |\sem{\varphi}(s) - \sem{\varphi}(t)| 
\eqpun .
\]

\subsection{Metrics for Concurrent Games}

We now extend the simulation and bisimulation metrics from
MDPs to general game structures.
As we shall see, unlike for MDPs, the a priori
and the a posteriori metrics do not coincide over games.
In particular, we show that the a priori formulation satisfies
both a tight logical characterization as well as reciprocity
while, perhaps surprisingly, the more natural a posteriori version does not.

A posteriori metrics are defined via the metric transformer 
$H_{\postsim_1}: \metrsp \mapsto \metrsp$ as follows, 
for all $d \in \metrsp$ and $s, t \in S$:
\begin{align}
  H_{\postsim_1}(d)(s,t) & = 
    \metr{s \loceq t} \imax 
    \Sup_{x_1 \in \dis_1(s)} \,
    \Inf_{y_1 \in \dis_1(t)} \,
    \Sup_{y_2 \in \dis_2(t)} \,
    \Inf_{x_2 \in \dis_2(s)} \,
    D(\trans(s,x_1,x_2),\trans(t,y_1,y_2),d) \nonumber \\[1ex]
  & = \metr{s \loceq t} \imax 
    \Sup_{x_1 \in \dis_1(s)} \,
    \Inf_{y_1 \in \dis_1(t)} \,
    \Sup_{y_2 \in \dis_2(t)} \,
    \Inf_{x_2 \in \dis_2(s)} \,
    \Sup_{k \in C(d)} \,
    \bigl(
    \E_{s}^{x_{1},x_{2}}(k) - \E_{t}^{y_{1},y_{2}}(k)
    \bigr) \eqpun .
    \label{eq-postsim}
\end{align}
A priori metrics are defined by bringing the $\sup_k$ outside. 
Precisely, we define a metric transformer 
$H_{\priosim_1}: \metrsp \mapsto \metrsp$ as follows, 
for all $d \in \metrsp$ and $s, t \in S$:
\begin{align}
  H_{\priosim_1}(d)(s,t) & =
    \metr{s \loceq t} \imax 
    \Sup_{k \in C(d)} \,
    \Sup_{x_1 \in \dis_1(s)} \,
    \Inf_{y_1 \in \dis_1(t)} \,
    \Sup_{y_2 \in \dis_2(t)} \,
    \Inf_{x_2 \in \dis_2(s)} \,
    \bigl(
    \E_{s}^{x_{1},x_{2}}(k) - \E_{t}^{y_{1},y_{2}}(k)
    \bigr) \nonumber \\[1ex]
  & = \metr{s \loceq t} \imax
    \Sup_{k \in C(d)} \Bigl[
    \Sup_{x_1 \in \dis_1(s)} \,
    \Inf_{x_2 \in \dis_2(s)} \E_{s}^{x_{1},x_{2}}(k) -
    \Sup_{y_1 \in \dis_1(t)} \,
    \Inf_{y_2 \in \dis_2(t)} \E_{t}^{y_{1},y_{2}}(k)
    \Bigr] \nonumber \\[1ex]
  & = \metr{s \loceq t} \imax 
    \Sup_{k \in C(d)} \bigl( 
    \pre_1(k)(s) - \pre_1(k)(t) \bigr) \eqpun .
    \label{eq-priosim}
\end{align}
First, we show that $H_{\priosim_1}$ and $H_{\postsim_1}$ are
monotonic in the lattice of metrics $(\metrsp, \leq)$.

\begin{lem}{} \label{lem-monotonicity}
The functions $H_{\priosim_1}$ and $H_{\postsim_1}$ are
monotonic in the lattice of metrics $(\metrsp, \leq)$.
\end{lem}

\begin{proof}
For $d, d' \in \metrsp$,
$d\leq d'$ implies $C(d) \subs C(d')$, 
and hence $\Sup_{k \in C(d)} (\pre_1(k)(s) - \pre_1(k)(t)) \le
\Sup_{k \in C(d')} (\pre_1(k)(s) - \pre_1(k)(t))$.
This shows the monotonicity of $H_{\priosim_1}$. 

The monotonicity of $H_{\postsim_1}$ can be shown in a similar fashion. 
From $d\leq d'$, reasoning as before we obtain 
\[
  \Sup_{k \in C(d)} \,
    \bigl(
    \E_{s}^{x_{1},x_{2}}(k) - \E_{t}^{y_{1},y_{2}}(k)
    \bigr)
  \leq
  \Sup_{k \in C(d')} \,
    \bigl(
    \E_{s}^{x_{1},x_{2}}(k) - \E_{t}^{y_{1},y_{2}}(k)
    \bigr) \eqpun .
\]
The result then follows from the monotonicity of the operators 
$\Sup_{x_1 \in \dis_1(s)}$, 
$\Inf_{y_1 \in \dis_1(t)}$, 
$\Sup_{y_2 \in \dis_2(t)}$,
$\Inf_{x_2 \in \dis_2(s)}$.
\end{proof}

On the basis of this lemma, we can define the least fixpoints of 
$H_{\priosim_1}$ and $H_{\postsim_1}$, which will yield our game
simulation and bisimulation metrics. 

\begin{defi}{} \label{def-metrics}
{\em A priori metrics:}
\begin{enumerate}[$\bullet$]
\item 
  The {\em a priori simulation metric} $[\priosim_1]$ is the least
  fixpoint of $H_{\priosim_1}$.
\item The {\em a priori bisimulation metric\/}
  $[\priobis_1]$ is the least symmetrical fixpoint of $H_{\priosim_1}$.
\end{enumerate}

\noindent {\em A posteriori metrics:}
\begin{enumerate}[$\bullet$]
\item 
  The {\em a posteriori game simulation metric\/} $[\postsim_1]$ is
  the least fixpoint of $H_{\postsim_1}$. 
\item 
  The {\em a posteriori game bisimulation metric\/} $[\postbis_1]$ is
  the least symmetrical fixpoint of $H_{\postsim_1}$. 
\end{enumerate}
By exchanging the roles of the players, we define the metric
transformers $H_{\priosim_2}$ and $H_{\postsim_2}$, and the metrics 
$[\priosim_2]$, $[\priobis_2]$, $[\postsim_2]$, $[\postbis_2]$. 
\end{defi}

We note that the a posteriori simulation metric $[\postsim_1]$ has
been introduced in \cite{luca-icalp-disc-03,rupak-thesis}.
We also note that the a posteriori bisimulation metric $[\postbis_i]$ can
be defined as the least fixpoint of $H_{\postbis_i}: \metrsp
\mapsto \metrsp$, defined for all $d \in \metrsp$ and 
$i \in \set{1,2}$ by 
\begin{equation}
  H_{\postbis_1}(d) = H_{\postsim_1}(d) \imax \text{\textit{Opp}} (H_{\postsim_1}(d)),
\end{equation}
where $\text{\textit{Opp}}(d) = \breve{d}$ denotes the opposite
of a metric $d$.
Similarly, the a priori bisimulation metric $[\priobis_i]$ can
be defined as the least fixpoint of $H_{\priobis_i}: \metrsp
\mapsto \metrsp$, defined for all $d \in \metrsp$ and 
$i \in \set{1,2}$ by 
\begin{equation}
  H_{\priobis_1}(d) = H_{\priosim_1}(d) \imax \text{\textit{Opp}} (H_{\priosim_1}(d)) \eqpun .
\end{equation}
We wish to show that the metrics of Definition~\ref{def-metrics} can
be computed via Picard iteration. 
To this end, it is necessary to show that the operators $H_{\postsim_1}$
and $H_{\priosim_1}$ on the lattice $(\metrsp, \leq)$ are upper semi-continuous. 
In fact, a very similar proof shows that the operators are lower
semi-continuous, and thus, continuous; we omit the proof of this more
general fact as it is not required for the desired result about the
applicability of Picard iteration. 

\begin{lem}{} \label{lem-continuity}
The operators $H_{\priosim_1}$
and $H_{\postsim_1}$ on the lattice $(\metrsp, \leq)$ are upper semi-continuous. 
\end{lem}

\begin{proof}
Let $D \subs \metrsp$ be an arbitrary set of distances, and let 
$d^* = \sup D$; note that $d^*$ exists, as $(\metrsp, \leq)$ is a
complete lattice.
 
We first prove the result for $H_{\priosim_1}$. 
We need to prove that 
$H_{\priosim_1}(\sup D) = \sup_{d \in D} H_{\priosim_1}(d)$, 
which we abbreviate 
$H_{\priosim_1}(\sup D) = \sup H_{\priosim_1}(D)$.
In one direction, 
$H_{\priosim_1}(\sup D) \geq \sup H_{\priosim_1}(D)$
follows from the monotonicity of $H_{\priosim_1}$
(Lemma~\ref{lem-monotonicity}). 
In the other direction, we will show that for all $\epsilon > 0$, 
there is $d \in D$ such that 
$|H_{\priosim_1}(d^*) - H_{\priosim_1}(d)| \leq \epsilon$, 
where for $d,d' \in \metrsp$, $|d-d'|$ is the 1-norm distance between
$d$ and $d'$.
For convenience, let $G(k) \in \metrsp$ be defined as 
$G(k)(s,t) = \pre_1(k)(s) - \pre_1(k)(t)$, so that 
we can write $H_{\priosim_1}(d) = \metr{s \loceq t} \imax \sup_{k \in C(d)} G(k)$. 

Given $\epsilon > 0$, choose $d \in D$ such that 
for all $s, t \in S$, we have  
$d(s,t) / d^*(s,t) \geq 1 - \epsilon / 4$ if $d^*(s,t) > 0$, and 
$d(s,t) = 0$ if $d^*(s,t) = 0$. 
Note that for all $k \in C(d^*)$, we have $(1 - \epsilon / 4) k \in C(d)$
and $|k - (1 - \epsilon / 4) k| \leq \epsilon / 4$, as $|k| \leq 1$.
Thus, $d \in D$ is such that 
for all $k \in C(d^*)$, there is $k' \in C(d)$ with 
$|k - k'| \leq \epsilon / 4$.
In other words, $d$ is such that the Hausdorff distance between
$C(d^*)$ and $C(d)$ is at most $\epsilon / 4$.
We now prove that for this $d$, we have
\begin{equation} \label{eq-in-proof-cont} 
|\sup_{k \in C(d^*)} G(k) - \sup_{k \in C(d)} G(k)| \leq \epsilon \eqpun . 
\end{equation}
In fact, let $k^* \in C(d^*)$ be such that 
\begin{equation} \label{eq-kstar-close}
 | G(k^*) - \sup_{k \in C(d^*)} G(k) | \leq \epsilon / 2 \eqpun . 
\end{equation}
and let $k' \in C(d)$ be such that $|k^* - k'| \leq \epsilon / 4$. 
For $s, t \in S$, 
we have by definition $G(k^*)(s,t) = \pre_1(k^*)(s) - \pre_1(k^*)(t)$; 
let 
\[
  x_1(s) = \arg \Sup_{x \in \dis_1(s)} \Inf_{y \in \dis_2(s)}
  \E^{x,y}_s (k^*) \eqpun .
\] 
By employing $x_1(s)$ at all $s \in S$, player~1 can guarantee 
\[
  | G(k')(s,t) - G(k^*)(s,t) | \leq \epsilon / 2,
\] 
which together with (\ref{eq-kstar-close}) leads to (\ref{eq-in-proof-cont}). 
In turn, (\ref{eq-in-proof-cont}) yields the result. 

We can prove the result for $H_{\postsim_1}$ following a similar
argument. 
Precisely, in one direction, 
$H_{\postsim_1}(\sup D) \geq \sup H_{\postsim_1}(D)$
follows from the monotonicity of $H_{\postsim_1}$
(Lemma~\ref{lem-monotonicity}). 
In the other direction, we will show that for all $\epsilon > 0$, 
there is $d \in D$ such that 
$|H_{\postsim_1}(d^*) - H_{\postsim_1}(d)| \leq \epsilon$, 
where for $d,d' \in \metrsp$, $|d-d'|$ is the 1-norm distance between
$d$ and $d'$.
Again, let $d$ be such that the Hausdorff distance between
$C(d^*)$ and $C(d)$ is at most $\epsilon/2$.
For such a $d$, we have that for all $s, t \in S$, and 
$x_1 \in \dis_1(s)$, 
$y_1 \in \dis_1(t)$, 
$x_2 \in \dis_2(s)$, 
$y_2 \in \dis_2(t)$, 
\[
  \Bigl| 
  \Sup_{k \in C(d^*)} \,
    \bigl(
    \E_{s}^{x_{1},x_{2}}(k) - \E_{t}^{y_{1},y_{2}}(k)
    \bigr) 
  -
  \Sup_{k \in C(d)} \,
    \bigl(
    \E_{s}^{x_{1},x_{2}}(k) - \E_{t}^{y_{1},y_{2}}(k)
    \bigr) 
  \Bigr| \leq \epsilon,
\]
and this leads easily to the result. 
\end{proof}


This result implies that we can compute $[\priosim_1]$ as the fixpoint 
of $H_{\priosim_1}$ via Picard iteration; we denote by 
$d_n = H_{\priosim_1}^n (\imeanbb{0})$ the $n$-iterate of this.
Similarly, we can compute 
$[\postsim_1]$ as the fixpoint of $H_{\postsim_1}$ via Picard iteration.

\begin{thm}{} \label{theo-picard}
The following assertions hold, for $i \in \set{1,2}$: 
\begin{enumerate}[\em(1)]
\item Let $d_0 = d'_0 = \imeanbb{0}$, and for $n \geq 0$, let 
\begin{align} \label{eq-picard-sim}
  d_{n+1} = H_{\priosim_i} (d_n) \quad \text{and} \quad
 d'_{n+1} = H_{\postsim_i}(d'_n) \eqpun . 
\end{align}
We have $\lim_{n \rightarrow \infty} d_n = [\priosim_i]$
and     $\lim_{n \rightarrow \infty}d'_n = [\postsim_i]$. 

\item Let $b_0 = b'_0 = \imeanbb{0}$, and for $n \geq 0$, let 
\begin{align} \label{eq-picard-bisim}
  & b_{n+1} = H_{\priosim_i}(b_n) \imax 
              \text{\textit{Opp}} (H_{\priosim_i}(b_n)) 
  \;\;\text{and}\;\;
%
  b'_{n+1} = H_{\postsim_i}(b'_n) \imax 
              \text{\textit{Opp}} (H_{\postsim_i}(b'_n)) \eqpun .
\end{align}
We have $\lim_{n \rightarrow \infty} b_n = [\priobis_i]$ and
$\lim_{n \rightarrow \infty} b'_n = [\postbis_i]$. 
\end{enumerate}
\end{thm}

\begin{proof}
The statements follow from the definitions of the metrics, and from 
Lemmas~\ref{lem-monotonicity} and~\ref{lem-continuity}. 
\end{proof}

We now show some basic properties of these metrics. First, we show that
the a priori
fixpoints give a (directed) metric, i.e., they are non-negative and
satisfy the triangle inequality. We also prove that the a priori
and a posteriori metrics are distinct. We then focus on the a priori 
metrics, and show, through our results, that they are the natural metrics 
for concurrent games.

\begin{thm}{}\label{theo-triangle}
For all game structures $\game$, and all states $s, t, u$ of $\game$,
we have,
\begin{enumerate}[\em(1)]
\item $[s \priosim_1 t] \geq 0$ and 
    $\metr{s \priosim_1 u} \le \metr{s \priosim_1 t} + \metr{t \priosim_1 u}.$
\item $[s \postsim_1 t] \geq 0$ and $\metr{s \postsim_1 u} \le \metr{s \postsim_1 t} + \metr{t \postsim_1 u}$.
\end{enumerate}
\end{thm}

\begin{proof}
We prove the following statement: if $d \in \metrsp$ is a directed metric,
then: 
\begin{enumerate}[(1)]
\item $H_{\priosim_1}(d)$ is a directed metric; 
\item $H_{\postsim_1}(d)$ is a directed metric.
\end{enumerate}
The theorem then follows by induction on the Picard iteration with
which the a priori and a posteriori metrics can be computed
(Theorem~\ref{theo-picard}). 
We prove the result first for the a priori metric. 

First, from $d' = H_{\priosim_1}(d)$ and $[\loceq] \geq 0$, we
immediately have $d' \geq 0$ (where inequalities are interpreted in
pointwise fashion). 

To prove the triangle inequality, we observe that   
$[s\loceq t] + [t\loceq u] \geq [s\loceq u]$ for all $s,t,u \in S$. 
Also, 
\begin{multline*}
\Sup_{k \in C(d)} \bigl(\pre_1(k)(s) - \pre_1(k)(t) \bigr) + 
\Sup_{k \in C(d)} \bigl(\pre_1(k)(t) - \pre_1(k)(u) \bigr) \\[1ex]  
\geq
\Sup_{k \in C(d)} \bigl( \pre_1(k)(s) - \pre_1(k)(t) + 
\pre_1(k)(t) - \pre_1(k)(u) \bigr) \\
= \Sup_{k \in C(d)} \bigl( \pre_1(k)(s) - \pre_1(k)(u) \bigr) \eqpun . 
\end{multline*}
Thus, we obtain 
\begin{multline*}
H_{\priosim_1}(d)(s,t) +  H_{\priosim_1}(d)(t,u) \\[1ex]
   = \bigl([s\loceq t] \imax 
        \Sup_{k \in C(d)}\bigl(
	 \pre_1(k)(s) - \pre_1(k)(t) \bigr)\bigr)
   +  \bigl([t\loceq u] \imax 
     \Sup_{k \in C(d)} \bigl(\pre_1(k)(t) - \pre_1(k)(u) \bigr)\bigr)
   \\[1ex]
   \geq \bigl([s\loceq u] \imax 
        \Sup_{k \in C(d)} \bigl( \pre_1(k)(s) -
        \pre_1(k)(u) \bigr) \bigr)
   = H_{\priosim_1}(d)(s,u),
\end{multline*}
leading to the result. 

For the a posteriori metric, let $d' = H_{\postsim_1}(d)$; again, we
can prove $d' \geq 0$ as in the a priori case. 
To prove the triangle inequality for $d'$, for $s,t \in S$, and for
distributions $x_1 \in \dis_1(s)$ and $y_1 \in \dis_1(t)$, it is
convenient to let 
\[
  G(x_1,y_1)(s,t) = 
  \Sup_{y_2 \in \dis_2(t)}
  \Inf_{x_2 \in \dis_2(s)} 
  \Sup_{k \in C(d)} 
  \bigl(\E_{s}^{x_{1},x_{2}}(k) - \E_{t}^{y_{1},y_{2}}(k) \bigr) ,
\]
With this notation, for $s,t,u \in S$, we have 
\begin{equation} \label{eq-triang-g}
  H_{\postsim_{1}}(d)(s,u) = [s \loceq u] \imax 
  \Sup_{x_1 \in \dis_1(s)}
  \Inf_{z_1 \in \dis_1(u)} G(x_1,z_1)(s,u) \eqpun . 
\end{equation}
Intuitively, the quantity $G(x_1,z_1)(s,u)$ 
is the distance between $s$ and $u$ computed in 
the 2-MDP obtained when player~1 plays $x_1$ at $s$ and $z_1$ at $u$. 
As a consequence of Theorem~\ref{thm-mpdpriovspost} (interpreted over
2-MDPs), and of the previous proof for the a-priori case, we have that
\begin{equation} \label{eq-triang-2mpds}
  G(x_1,z_1)(s,u) 
  \leq G(x_1,y_1)(s,t) + G(y_1,z_1)(t,u) \eqpun .
\end{equation}
for all $x_1 \in \dis_1(s)$, $y_1 \in \dis_1(t)$, and $z_1 \in \dis_1(u)$. 
This observation will be useful in the following. 

For any $\epsilon > 0$, let $x_1^*$ realize the $\sup$ in
(\ref{eq-triang-g}) within $\epsilon$, that is, 
\begin{equation} \label{eq-triang-g-2}
  \Inf_{z_1 \in \dis_1(u)} G(x^*_1,z_1)(s,u) \geq 
  \Sup_{x_1 \in \dis_1(s)}
  \Inf_{z_1 \in \dis_1(u)} G(x_1,z_1)(s,u) - \epsilon , 
\end{equation}
and let $z^*_1$ realize the $\Inf$ of the left-hand side of
(\ref{eq-triang-g-2}) also within $\epsilon$.
Intuitively, $x^*_1$ is the player-1 distribution at $s$ that is
hardest to imitate from $u$, and $z^*_1$ is the best imitation
of $x^*_1$ available at $u$.  
In the same fashion, let $y^*_1$ realize the $\inf$ within 
$\epsilon$ in 
$\Inf_{y_1 \in \dis_1(t)} G(x^*_1,y_1)(s,t)$, 
and let $z'_1$ realize the $\inf$ within 
$\epsilon$ in 
$\Inf_{z_1 \in \dis_1(u)} G(y^*_1,z_1)(t,u)$. 
In intuitive terms, 
$y^*_1$ is the imitator of $x^*_1$ in $t$, and 
$z'_1$ is the imitator of $y^*_1$ in $u$. 

We consider two cases. 
If $[s \loceq u] = 1$, then we are sure that the triangle inequality 
\begin{equation} \label{eq-triangle-post}
d'(s,u) \leq d'(s,t) + d'(t,u),
\end{equation}
holds. 
Otherwise, note that
\begin{equation} \label{eq-dbound-1}
  d'(s,u) \leq G(x^*_1,z^*_1)(s,u) + 2 \epsilon \eqpun .
\end{equation}
Since $x^*_1$ is not necessarily the distribution at $s$ that is
hardest to imitate from $t$, and since $y^*_1$ is not necessarily the
distribution at $t$ that is hardest to imitate from $u$, we also have: 
\begin{equation} \label{eq-dbound-2}
  d'(s,t) \geq G(x^*_1,y^*_1)(s,t) - \epsilon 
  \qquad \qquad
  d'(t,u) \geq G(y^*_1,z'_1)(t,u) - \epsilon \eqpun . 
\end{equation}
Since the triangle inequality holds for MDPs, as stated by
(\ref{eq-triang-2mpds}), we have 
\begin{equation} \label{eq-triang-mdps-bound} 
  G(x^*_1,z'_1)(s,u) 
  \quad \leq \quad G(x^*_1,y^*_1)(s,t) + G(y^*_1,z'_1)(t,u)
  \quad \leq \quad d'(s,t) + d'(t,u) + 2 \epsilon \eqpun . 
\end{equation}
Since $z^*_1$ is the best imitator of $x^*_1$ at $u$, we also have
\begin{equation} \label{eq-dbound-3}
  G(x^*_1,z^*_1)(s,u) - \epsilon \leq G(x^*_1,z'_1)(s,u),
\end{equation}
which together with (\ref{eq-triang-mdps-bound}) yields
\begin{equation} \label{eq-dbound-4}
  G(x^*_1,z^*_1)(s,u) \leq d'(s,t) + d'(t,u) + 3 \epsilon \eqpun . 
\end{equation}
From the choice of $x^*_1$, this finally leads to 
\[
  d'(s,u) \leq d'(s,t) + d'(t,u) + 5 \epsilon,
\]
for all $\epsilon > 0$, which yields the desired triangle inequality 
(\ref{eq-triangle-post}). 
\end{proof}

\paragraph{{\em A priori} and {\em a posteriori} metrics are distinct.}

First, we show that a priori and a posteriori metrics are distinct in
general: the a priori metric never exceeds the a posteriori one, and
there are concurrent games where it is strictly smaller. 
Intuitively, this can be explained as follows.  
Simulation entails trying to simulate the expectation of a valuation
$k$, as we see from (\ref{eq-postsim}), (\ref{eq-priosim}). 
It is easier to simulate a state $s$ from a state $t$ if the valuation
is known in advance, as in a priori metrics (\ref{eq-priosim}), than
if the valuation $k$ is chosen after all the moves have been chosen,
as in a posteriori metrics (\ref{eq-postsim}). 

As a special case, we shall see that equality holds for turn-based
game structures, in addition to MDPs as we have seen in the previous
subsection. 

\begin{thm}{} \label{theo-metrics-diff} The following assertions hold. 
\begin{enumerate}[\em(1)]
\item
For all game structures $\game$, and for all states $s, t$ of $\game$,
we have $[s \priosim_1 t] \leq [s \postsim_1 t]$. 

\item
There is a game structure $\game$, and states $s, t$ of $\game$, 
such that $[s \priosim_1 t] = 0$ and $[s \postsim_1 t] > 0$. 

\item
For all turn-based game structures,
we have $[{\priosim_1}] = [{\postsim_1}]$.

\end{enumerate}
\end{thm}

\begin{figure}\centering
\begin{minipage}[b]{0.29\linewidth}
\begin{tabular}{| r || c | c |}
\hline 
$\trans(t,*,*)(w)$ & $f$ & $g$ \\ \hline \hline
$b$ & $1/9$ & $5/9$ \\
$c$ & $4/9$ & $8/9$ \\
\hline
\end{tabular}
\end{minipage}
\hspace{0.5cm}
\begin{minipage}[b]{0.29\linewidth}
\begin{tabular}{| r || c | c |}
\hline 
$\trans(t,*,*)(u)$ & $f$ & $g$ \\ \hline \hline
$b$ & $8/9$ & $4/9$ \\
$c$ & $5/9$ & $1/9$ \\
\hline
\end{tabular}
\end{minipage}

\vspace{16pt}

\begin{minipage}[b]{0.29\linewidth}
\begin{tabular}{| r || c | c |}
\hline 
$\trans(s,*,*)(w)$ & $f$ & $g$ \\ \hline \hline
$a$ & $1/3$ & $2/3$ \\
\hline
\end{tabular}
\end{minipage}
\hspace{0.5cm}
\begin{minipage}[b]{0.29\linewidth}
\begin{tabular}{| r || c | c |}
\hline 
$\trans(s,*,*)(u)$ & $f$ & $g$ \\ \hline \hline
$a$ & $2/3$ & $1/3$ \\
\hline
\end{tabular}
\end{minipage}

\vspace{16pt}

\includegraphics[height=1.2in,width=4in]{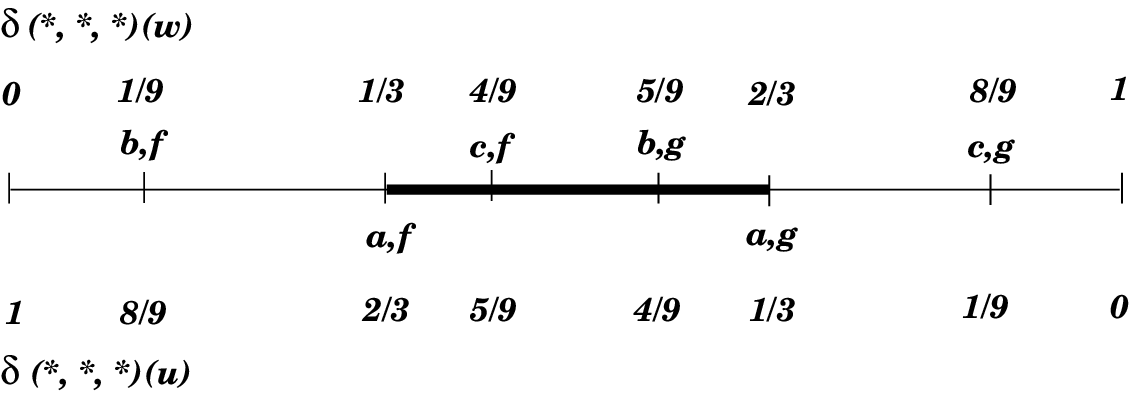}
\caption{A game that shows that the a priori and the a posteriori
metrics may not coincide.
The tables above show the transition probabilities from 
states $t$ and $s$ to states $w$ and $u$ for pure moves of the two
players. 
The row player is player~1 and the column player is player~2.
The line below is the two dimensional probability simplex 
that shows the transition probabilities induced by convex 
combinations of pure moves of the two players.
}
\label{fig-apriori-aposteriori}
\end{figure}

\begin{proof}
The first assertion is a consequence of the fact that, for all functions 
$f: \reals^2 \mapsto \reals$, we have $\sup_x \inf_y f(x,y) \leq
\inf_y \sup_x f(x,y)$. 
By repeated applications of this, we can show that, for all $d \in
\metrsp$, we have $H_\priosim(d) \leq H_\postsim(d)$ (with pointwise
ordering).
The result then follows from the monotonicity of $H_\priosim$ and
$H_\postsim$. 

For the second assertion, we give an example where a priori distances are
strictly less than a posteriori distances.
Consider a game with states $S = \set{s, t, u, w}$. 
States $u$ and $w$ are sink states with $[u \loceq w] = 1$; 
states $s$ and $t$ are such that $[s \loceq t] = 0$. 
At states $s$ and $t$, player~2 has moves $\{f, g\}$. 
Player-1 has a single move $\{ a \}$ at 
state $s$, and moves $\{b, c \}$ at state $t$. 
The moves from $s$ and $t$ lead to $u$ and $w$ with transition
probabilities indicated in Figure~\ref{fig-apriori-aposteriori}.
In the figure, the point $b,f$ indicates the probability of going to
$u$ and $w$ when the move pair $(b,f)$ is played, with 
$\trans(s,b,f)(u) + \trans(s,b,f)(w) = 1$; similarly for the other
move pairs. 
The thick line segment between the points 
$a,f$ and $a,g$ represents the transition probabilities arising when
player~1 plays move $a$, and player~2 plays a mixed move (a mix of $f$
and $g$). 

We show that, in this game, we have $[s \postsim_{1} t] > 0$. 
Consider the metric $d$ where $d(u,w) = 1$ 
(recall that $[u \loceq w] = 1$, and note the other distances do 
not matter, since $u$, $w$ are the only two destinations). 
We need to show 
\begin{align}
  \forall {y_1 \in \dis_1(t)} . 
  \exists {y_2 \in \dis_2(t)} . 
  \forall {x_2 \in \dis_2(s)} .
  \exists {k \in C(d)} .
  \bigl(
  \E_{s}^{a,x_{2}}(k) - \E_{t}^{y_{1},y_{2}}(k)
  \bigr) > 0 \eqpun . 
  \label{eq-popout}
\end{align}
Consider any mixed move $y_1 = \alpha b + (1-\alpha)c$, where $b, c$
are the moves available to player~1 at $t$, and $0 \leq \alpha \leq 1$. 
If $\alpha \geq \frac{1}{2}$, choose move $f$ from $t$ as $y_2$,
    and choose $k(w) = 1$, $k(u) = 0$. 
Otherwise, choose move $g$ from $t$ as $y_2$, 
    and choose $k(w) = 0$, $k(u) = 1$. 
With these choices, the transition probability $\trans(t,y_1,y_2)$
will fall outside of the segment $[(a,f), (a,g)]$ in
Figure~\ref{fig-apriori-aposteriori}. 
Thus, with the choice of $k$ above, we ensure that the
difference in (\ref{eq-popout}) is always positive. 

To show that in the game we have $[s \priosim_{1} t] = 0$, 
it suffices to show (given that $[s\priosim_{1} t] \geq 0$) that 
\begin{align*}
  \forall k \in C(d) . 
  \exists y_1 \in \dis_1(t) . 
  \forall y_2 \in \dis_2(t) . 
  \exists x_2 \in \dis_2(s) .
  \bigl(
  \E_{s}^{a,x_{2}}(k) - \E_{t}^{y_{1},y_{2}}(k)
  \bigr) \leq 0 \eqpun . 
\end{align*}
If $k(u) = k(w)$, the result is immediate. 
Assume otherwise, that $k(u) < k(w)$, 
and choose $y_1 = c$.
For every $y_2$, the distribution over successor states (and of
$k$-expectations) will be in the interval $[(c,f), (c,g)]$ in
Figure~\ref{fig-apriori-aposteriori}. 
By choosing $x_2 = f$, we have that 
$\E_s^{a,f}(k) < \E_t^{c,y_2}(k)$ for all $y_2 \in \dis_2(t)$, leading
to the result.
Similarly if $k(u) > k(w)$, by choosing $y_1 = b$, the distribution 
over successor states (and of $k$-expectations) will now be in the
interval $[(b,f), (b,g)]$.
By choosing $x_2 = g$, we have that
$\E_s^{a,g}(k) < \E_t^{b,y_2}(k)$ for all $y_2 \in \dis_2(t)$, 
again leading to the result.

The last assertion of the theorem is proved in the same way as 
Theorem~\ref{thm-mpdpriovspost}.  
\end{proof}

\paragraph{Reciprocity of {\em a priori} metric.}

The previous theorem establishes that the a priori and a posteriori
metrics are in general distinct. 
We now prove that it is the a priori metric, rather than
the a posteriori one, that enjoys reciprocity, and that provides a
(quantitative) logical characterization of $\qmu$. 
We begin by considering reciprocity. 

\begin{thm}{} \label{theo-recipro-prio}
The following assertions hold. 
\begin{enumerate}[\em(1)]
\item For all game structures $\game$, we have 
$[{\priosim_1}] = [{\priomis_2}]$, and 
$[{\priobis_1}] = [{\priobis_2}]$. 

\item There is a concurrent game structure $\game$, with states $s$
and $t$, where $[\postsim_1] \neq [\postmis_2]$. 

\item There is a concurrent game structure $\game$, with states $s$
and $t$, where $[\postbis_1] \neq [\postbis_2]$. 

\end{enumerate}
\end{thm}

\begin{proof}
For the first assertion, it suffices to show that, for all 
$d \in \metrsp$, and states $s, t \in S$, we have 
$H_{\priosim_1}(d)(s,t) = H_{\priosim_2}(\breve{d})(t,s)$. 
We proceed as follows: 
\begin{align}
  & \Sup_{k \in C(d)}
    \bigl(\pre_1(k)(s) - \pre_1(k)(t)\bigr) \label{apriorec1} \\
= & \Sup_{k \in C(d)}
    \bigl(- \pre_2(1 - k)(s) + \pre_2(1 - k)(t)\bigr) \label{apriorec2} \\
= & \Sup_{k \in C(\breve{d})}
    \bigl(\pre_2(k)(t) - \pre_2(k)(s)\bigr) \eqpun . \label{apriorec3} 
\end{align}
The step from (\ref{apriorec1}) to (\ref{apriorec2}) uses 
$\pre_1(k)(s) = 1 - \pre_2 (1 - k)(s)$ \cite{vonNeumannMorgenstern44,dAM04}, 
and the step from (\ref{apriorec2}) to (\ref{apriorec3}) uses
the change of variables $k \go 1 - k$. 

For the second assertion, 
consider again the game of Figure~\ref{fig-apriori-aposteriori}. 
We will show that $[t \postsim_{2} s] = 0$.
Together with $[s \postsim_1 t] > 0$, as shown in the proof of 
Theorem~\ref{theo-metrics-diff}, this leads to the result. 
To obtain the result, we will prove that for all $d$, we have: 
\[
  \forall {y_2 \in \dis_2(t)} . 
  \exists {x_2 \in \dis_2(s)} . 
  \exists {y_1 \in \dis_1(t)} . 
  \forall {k \in C(d)} .
  \bigl(
  \E_{t}^{y_2,y_1}(k) - \E_{s}^{x_2,a}(k)
  \bigr) = 0 \eqpun . 
\]
where we have used the fact that player~1 at $s$ plays $x_1 = a$.
Any mixed move $y_2 \in \dis_2(t)$ can be written as 
$y_2 = \alpha f + (1-\alpha)g$ for $0 \leq \alpha \leq 1$.
Choose $y_1 = \alpha c + (1-\alpha)b$, and 
\[
   x_2 = \alpha \Bigl(\frac{2}{3}f + \frac{1}{3}g \Bigr) + 
   (1-\alpha) \Bigl( \frac{1}{3}f + \frac{2}{3}g \Bigr) \eqpun .
\]
Under this choice of mixed moves, we have: 
\begin{align*}
  \trans(t,y_1,y_2)(w) & = \frac{4}{9} \alpha^2 + \alpha (1-\alpha) 
                  + \frac{5}{9} (1-\alpha)^2 
                = \frac{5}{9} - \frac{1}{9}\alpha \\[1ex]
  \trans(s,x_1,x_2)(w) & = \alpha \Bigl( 
                        \frac{2}{3} \cdot \frac{1}{3} + 
                        \frac{1}{3} \cdot \frac{2}{3} \Bigr) 
                  + (1-\alpha) \Bigl( 
                        \frac{2}{3} \cdot \frac{2}{3} + 
                        \frac{1}{3} \cdot \frac{1}{3} \Bigr) 
                = \frac{5}{9} - \frac{1}{9} \alpha \eqpun . 
\end{align*}
As the probabilities of transitions to $w$ are equal from $t$ and $s$,
we obtain that for all ${k \in C(d)}$, we have 
$\E_{t}^{y_2,y_1}(k) - \E_{s}^{x_2,a}(k) = 0$, 
as desired. 

For the third assertion, we consider a modified version of the game
depicted in Figure~\ref{fig-apriori-aposteriori}, obtained by adding
two new moves to player~2 at state $t$, namely $f'$ and $g'$. 
We define the transition probabilities of these new moves by 
\[
  \trans(t, \ast, f') = \trans (s, a, f) 
  \qquad 
  \trans(t, \ast, g') = \trans (s, a, g) \eqpun . 
\]
To prove $[s \postsim_1 t] > 0$, we can proceed as in the proof of 
Theorem~\ref{theo-metrics-diff}, noting that we can choose $y_2$ as in
that proof (this is possible, as player~2 at $t$ has \emph{more} moves
available in the modified game). 
This leads to $[s \postbis_1 t] > 0$. 

To show that $[s \postbis_2 t] = 0$, given the transition structure of
the game, it suffices to show that $[s \postsim_2 t] = 0$ and 
$[t \postsim_2 s] = 0$. 
To show that $[s \postsim_2 t] = 0$, we show that for all $d$, we have:
\[
  \forall x_2 \in \dis_2(s) . 
  \exists y_2 \in \dis_2(t) . 
  \forall y_1 \in \dis_1(t) . 
  \forall {k \in C(d)} .
  \bigl(
  \E_{s}^{x_2,a}(k) - \E_{t}^{y_2,y_1}(k)
  \bigr) = 0 \eqpun . 
\]
We can write any mixed move $x_2 \in \dis_2(s)$ as
$x_2 = \alpha f + (1 - \alpha) g$. 
We can then choose $y_2 = \alpha f' + (1 - \alpha) g'$, 
and since at $t$ under $f'$, $g'$ the transition probabilities do not depend
on the mixed move $y_1$ chosen by player~1, we have that the transition
probabilities from $s$ and $t$ match for all $0 \leq \alpha \leq 1$. 

To show that $[t \postsim_2 s] = 0$, we need to show that: 
\[
  \forall y_2 \in \dis_2(t) . 
  \exists x_2 \in \dis_2(s) . 
  \exists y_1 \in \dis_1(t) . 
  \forall {k \in C(d)} .
  \bigl(
  \E_{t}^{y_2,y_1}(k) - \E_{s}^{x_2,a}(k)
  \bigr) = 0 \eqpun . 
\]
Any mixed move $y_2 \in \dis_2(t)$ can be written as 
\[
  y_2 = \gamma \Bigl[\alpha f + (1 - \alpha) g \Bigr] 
        + (1 - \gamma) \Bigl[ \beta f' + (1 - \beta) g' \Bigr] , 
\]
for some $\alpha, \beta, \gamma \in [0,1]$. 
We choose $x_2$ and $y_1$ as follows: 
\begin{align*}
  x_2 & = \alpha \gamma \Bigl[ \frac{2}{3}f + \frac{1}{3}g \Bigr]
        + (1 - \alpha) \gamma \Bigl[ \frac{1}{3} f + \frac{2}{3} g \Bigr]
        + (1 - \gamma) \Bigl[ \beta f + (1 - \beta) g \Bigr] \\[1ex]
  y_1 & = \alpha c + (1 - \alpha) b \eqpun .
\end{align*}
With these mixed moves, we have 
$\trans(s,a,x_2) = \trans(t,y_1,y_2)$, leading to the result. 
\end{proof}

\medskip 
\noindent
As a consequence of this theorem, we write $[\priobis_g]$ in place of
$[\priobis_1] = [\priobis_2]$, to emphasize that the player~1 and
player~2 versions of game equivalence metrics coincide.

\paragraph{Logical characterization of {\em a priori} metric.}

We now prove that $\qmu$ provides a logical characterization for
the a priori metrics. We first state and prove two lemmas
that lead to the desired result. The proof of the lemmas use ideas from 
\cite{rupak-thesis} and \cite{DGJP02}. We recall from 
Theorem~\ref{theo-picard} that we can compute $[\priosim_1]$ via 
Picard iteration, with $d_n = H_{\priosim_1}^n (\imeanbb{0})$ being
the $n$-iterate. 

We prove the existence of a logical characterization via a sequence of
the following two lemmas. 
The first lemma proves that a priori metrics provide a bound for the
difference in value of $\qmu$-formulas. 

\begin{lem}{} \label{val-le-metric}
The following assertions hold for all game structures.
\begin{enumerate}[\em(1)]
\item 
For all $\varphi \in \qmu_1^+$, and for all $s, t \in S$, we have 
$\sem{\varphi}(s) - \sem{\varphi}(t) \leq \metr{s \priosim_{1} t}$.
\item 
For all $\varphi \in \qmu$, and for all $s, t \in S$, we have  
$\vert \sem{\varphi}(s) - \sem{\varphi}(t) \vert \le \metr{s \priobis_g t}$.
\end{enumerate}
\end{lem}

\begin{proof}
We prove the first assertion. 
The proof is by induction on the structure of a (possibly open) formula $\varphi \in \qmu_1^+$.
Call a variable valuation $\xenv$ {\em bounded} if,
for all variables $\mvar\in\mvars$ and states $s,t$, we have that
$\xenv(\mvar)(s) - \xenv(\mvar)(t) \le \metr{s\priosim_1 t}$.
We prove by induction that 
for all $s, t \in S$, for all bounded variable valuations $\xenv$, 
we have $\sem{\varphi}_\xenv(s) - \sem{\varphi}_\xenv(t) \le \metr{ s \priosim_1 t}$.
For clarity, we sometimes omit writing the variable valuation $\xenv$.

The base case for constants is trivial, and the case
for observation variables follows since $\metr{s \equiv t} \le \metr{ s \priosim_1 t}$. 
The case for variables $\mvar \in\mvars$ follows from the assumption of bounded variable valuations.
For $\varphi_1 \vee\varphi_2$, assume the induction hypothesis for $\varphi_1$, $\varphi_2$, 
and note that 
\begin{multline*}
         \bigl( \sem{\varphi_1}(s) \imax \sem{\varphi_2}(s) \bigr) - 
          \bigl( \sem{\varphi_1}(t) \imax \sem{\varphi_2}(t) \bigr) \\[1ex]
\leq  \bigl( \sem{\varphi_1}(s) - \sem{\varphi_1}(t) \bigr) \imax 
          \bigl( \sem{\varphi_2}(s) - \sem{\varphi_2}(t) \bigr)
          \leq \metr{ s \priosim_1 t} \eqpun .
\end{multline*}
The proof for $\wedge$ is similar.
For $\varphi_1 \oplus c$ and $\varphi_1 \ominus c$, we have by induction hypothesis
that $\sem{\varphi_1}(s) - \sem{\varphi_1}(t) \le \metr{s \priosim_1 t}$, and so
the ``shifted versions'' also satisfy the same bound. 

For the induction step for $\qpre_1$, assume the induction hypothesis
for $\varphi$, and note that we can choose $k \in C(\metr{\priosim_1})$ such that 
$k(s) = \sem{\varphi}(s)$ at all $s \in S$. 
We have, for all $s, t \in S$, 
\begin{equation} \label{eq-pre-induction-sem}
  \sem{\qpre_1(\varphi)}(s) - \sem{\qpre_1(\varphi)}(t)
  \leq \sup_{k \in C(\metr{\priosim_1})} \bigl( \pre_1(k)(s) - \pre_1(k)(t) \bigr) 
  \leq \metr{s \priosim_1 t} \eqpun . 
\end{equation}
where the last inequality follows by noting that $\metr{\priosim_1}$ is a fixpoint
of $H_{\priosim_1}$.

The proof for the fixpoint operators is performed by considering their
Picard iterates.
We consider the case $\mu Z.\varphi$, 
the proof for $\nu Z.\varphi$ is similar. 
Let $\xenv$ be a bounded variable valuation.
Then, the variable valuation $\xenv_0 = \xenv[Z\mapsto \imeanbb{0}]$ is also 
bounded, and by induction hypothesis, the formula $\varphi$ when 
evaluated in the variable valuation $\xenv_0$ satisfies
\begin{equation}\label{eq-bounded-variable-valuation}
\sem{\varphi}_{\xenv_0}(s) - \sem{\varphi}_{\xenv_0}(t) \le \metr{s\priosim_1 t} \eqpun .
\end{equation}
Now consider the variable valuation $\xenv_1 = \xenv[Z\mapsto \sem{\varphi}_{\xenv_0}]$.
From Equation~\eqref{eq-bounded-variable-valuation}, we get that $\xenv_1$ is bounded,
and again, by induction hypothesis, we have that 
$\sem{\varphi}_{\xenv_1}(s) - \sem{\varphi}_{\xenv_1}(t) \le \metr{s\priosim_1 t}$.
In general, for $k\ge 0$, consider the variable valuation  $\xenv_{k+1} = \xenv[Z\mapsto \sem{\varphi}_{\xenv_k}]$.
By the above argument, each variable valuation $\xenv_k$ is bounded, and so for every $k\ge 0$, we have
\begin{equation}\label{eq-bounded-variable-valuation-k}
\sem{\varphi}_{\xenv_k}(s) - \sem{\varphi}_{\xenv_k}(t) \le \metr{s\priosim_1 t} \eqpun .
\end{equation}
Taking the limit, as $k\rightarrow \infty$, 
we have that 
\begin{equation}\label{eq-bounded-variable-valuation-limit}
\lim_{k\rightarrow\infty} (\sem{\varphi}_{\xenv_k}(s) - \sem{\varphi}_{\xenv_k}(t)) =
\sem{\mu Z.\varphi}_\xenv(s) - \sem{\mu Z.\varphi}_\xenv(t) \le \metr{s\priosim_1 t} \eqpun .
\end{equation}

The proof of the second assertion can be done along the same lines, using
the symmetry of $\priobis_g$. 
The proof is again by induction on the structure of the formula.
In particular, (\ref{eq-pre-induction-sem}) can be proved for either
player: for $n \geq 0$ and $i \in \set{1,2}$, 
\[
  \sem{\qpre_i(\varphi)}(s) - \sem{\qpre_i(\varphi)}(t)
  \leq \sup_{k \in C(\metr{\priobis_g})} \bigl( \pre_i(k)(s) - \pre_i(k)(t) \bigr) 
  \leq \metr{s \priobis_g t} \eqpun .
\]
Negation can be dealt with by noting that 
$
  \sem{\neg\varphi}(s) - \sem{\neg\varphi(t)} = 
  \sem{\varphi}(t) - \sem{\varphi(s)}
$, 
and by using the symmetry of $\priobis_g$; the other cases
are similar. 
\end{proof}

The second lemma states that the $\qmu$ formulas can attain the distance
computed by the simulation metric. 

\begin{lem}{} \label{metric-le-val}
The following assertions hold for all game structures $\game$, and for all
states $s, t$ of $\game$.
\begin{align*}
  \metr{s \priosim_{1} t} 
  & \leq \sup_{\varphi \in \qmu_1^+} (\sem{\varphi}(s) - \sem{\varphi}(t)) \\[1ex]
  \metr{s \priobis_{g} t} 
  & \leq \sup_{\varphi \in \qmu} |\sem{\varphi}(s) - \sem{\varphi}(t)|
\end{align*}
\end{lem}

\begin{proof}
We show by induction on $n$ that $d_{n}(s, t) \le \Sup_{\varphi \in \qmu}  
(\sem{\varphi}(s) - \sem{\varphi}(t))$. 
The base case is trivial. 
For the induction step, the distance is: 
\begin{equation} \label{d-i+1-def}
  d_{i+1}(s, t) = \Sup_{k \in C(d_{i})} 
  \bigl(\pre_1(k)(s) - \pre_1(k)(t)\bigr) \eqpun .
\end{equation}
The challenge is to show that, for all $s, t \in S$, we can construct
a formula $\psi_{st}$ that witnesses the distance within an arbitrary
$\ve > 0$:  
\begin{equation} \label{prop-of-psi}
  d_{i+1}(s, t) - \ve \leq \sem{\psi_{st}}(s) - \sem{\psi_{st}}(t) \eqpun . 
\end{equation}
To this end, let $\kstar$ be the value of $k$ that realizes the $\Sup$ in 
(\ref{d-i+1-def}) within $\ve/4$.
By induction hypothesis, for each pair of states $s'$ and $t'$
we can choose $\varphi'_{s't'}$ such that
\begin{equation} \label{writeup1}
  d_{i}(s', t') - \ve/4 \le \sem{\varphi'_{s't'}}(s') - \sem{\varphi'_{s't'}}(t') \eqpun . 
\end{equation}
Let $\varphi_{s't'}$ be a shifted version of $\varphi'_{s't'}$, such
that $\varphi_{s't'}(s') = \kstar(s')$:
\begin{equation} \label{eq-shift}
  \varphi_{s't'} = \varphi'_{s't'} \oplus (\kstar(s') - \sem{\varphi'_{s't'}}(s')) \eqpun .
\end{equation}
We now prove that: 
\begin{align}
  \sem{\varphi_{s't'}}(s') & = \kstar(s') \label{write-rel1} \\[1ex]
  \sem{\varphi_{s't'}}(t') & \leq \kstar(t') + \ve/4 \eqpun . \label{write-rel2}
\end{align}
Equality (\ref{write-rel1}) is immediate from (\ref{eq-shift}). 
We prove (\ref{write-rel2}) as follows. 
We can rewrite (\ref{writeup1}) as 
\begin{equation} \label{writeup1-bis}
  \sem{\varphi'_{s't'}}(t') - \ve/4 \leq \sem{\varphi'_{s't'}}(s') - d_{i}(s', t') \eqpun . 
\end{equation}
Since $\kstar \in C(d_i)$, we have $\kstar(s') - \kstar(t') \leq d_i(s',t')$,
or
\begin{equation} \label{kstarbound}
  \kstar(t') - \kstar(s') \geq - d_i (s', t') \eqpun . 
\end{equation}
Plugging this relation into (\ref{writeup1-bis}), we obtain 
\begin{equation}
  \sem{\varphi'_{s't'}}(t') - \ve/4 \leq \sem{\varphi'_{s't'}}(s') + 
\kstar(t') - \kstar(s') \eqpun .
\end{equation}
Plugging this relation into (\ref{eq-shift}) evaluated at $t'$, we
obtain 
\[
  \sem{\varphi_{s't'}}(t') - \ve/4 \hspace{0.7em} \leq \hspace{0.7em}   \sem{\varphi'_{s't'}}(s') + 
  \kstar(t') - \kstar(s')
   \oplus \bigl( \kstar(s') - \sem{\varphi'_{s't'}}(s') \bigr),
\]
or
\[
  \sem{\varphi_{s't'}}(t') - \ve/4 \hspace{0.9em} \leq \hspace{0.9em} \kstar(t')  - \bigl( \kstar(s') - 
  \sem{\varphi'_{s't'}}(s')\bigr)
          \oplus \bigl( \kstar(s') - \sem{\varphi'_{s't'}}(s')\bigr) 
          \hspace{0.9em} \leq \hspace{0.9em} \kstar(t'),
\]
which proves (\ref{write-rel2}). 
Define now 
$
  \textstyle  \varphi_{s'} = \bigwedge_{t'} \varphi_{s't'}. 
$
From (\ref{write-rel1}) and (\ref{write-rel2}) we have 
\begin{align}
  \sem{\varphi_{s'}}(s') & = \kstar(s') \label{bwrite-rel1} \\[1ex]
  \sem{\varphi_{s'}}(t') & \leq \kstar(t') + \ve/4 \eqpun . \label{bwrite-rel2}
\end{align}
Define then 
$
  \textstyle  \varphi = \bigvee_{s'} \varphi_{s'}. 
$
From (\ref{bwrite-rel1}), (\ref{bwrite-rel2}), we have that 
\begin{equation} \label{phi-is-k}
  \kstar(s') \leq \sem{\varphi}(s') \leq \kstar(s') + \ve/4 \eqpun .
\end{equation}
for all $s' \in S$. 
As formula $\psi_{st}$, we propose thus to take the formula $\qpre(\varphi)$. 
From (\ref{phi-is-k}), we have that 
$|\sem{\psi_{st}}(s) - \pre_1(\kstar)(s)| \leq \ve/4$, and similarly, 
$|\sem{\psi_{st}}(t) - \pre_1(\kstar)(t)| \leq \ve/4$.
By comparison with (\ref{d-i+1-def}), and by the fact that $\kstar$
realizes the $\sup$ within $\ve/4$, we finally have
(\ref{prop-of-psi}), as desired. 
\end{proof}

From these two lemmas, we can conclude that $\sem{\qmu}$ provides a
logical characterization for the a priori metrics, as stated by the
next theorem.

\begin{thm}{} \label{theo-logical-charact-metrics}
The following assertions hold for all game structures $\game$, and for all
states $s, t$ of $\game$: 
\begin{align*}
  \metr{s \priosim_{1} t} 
  & = \sup_{\varphi \in \qmu_1^+} (\sem{\varphi}(s) - \sem{\varphi}(t)) &
  \metr{s \priobis_{g} t} 
  & = \sup_{\varphi \in \qmu} |\sem{\varphi}(s) - \sem{\varphi}(t)|
\end{align*}
\end{thm}

We note that, due to Theorem~\ref{theo-metrics-diff}, an analogous
result does not hold for the a posteriori metrics. 
Together with the lack of reciprocity of the a posteriori metrics,
this is a strong indication that the a priori metrics, and not the a
posteriori ones, are the ``natural'' metrics on concurrent games. 

Our metrics are not characterized by
the probabilistic temporal logic PCTL 
\cite{HanssonJ94,BerkP95}. 
In fact, the values of PCTL formulas can change from true to false 
when certain probabilities cross given thresholds, so that PCTL 
formulas can have different boolean values on games that are very 
close in transition probabilities, and hence, very close in our metric.
Quantitative metrics such as the ones developed in this paper are 
suited to quantitative-valued formulas, such as those of $\qmu$.

\subsection{The Kernel}

The kernel of the metric $[\priobis_g]$ defines an equivalence relation
$\priobis_g$ on the states of a game structure: $s \priobis_g t$ iff $[s
\priobis_g t] = 0$.
We call this the {\em game bisimulation\/} relation.
Notice that by the reciprocity property of $\priobis_g$, the game
bisimulation relation is canonical: ${\priobis_1} = {\priobis_2} =
{\priobis_g}$.
Similarly, we define the {\em game simulation\/} preorder $s \priosim_1
t$ as the kernel of the directed metric $[\priosim_1]$, that is,
$s\priosim_1 t$ iff $[s \priosim_1 t] = 0$.
%
%
Alternatively, it is possible to define $\priosim_1$ and $\priobis_g$
directly. 
Given a relation $R \subs S \times S$, let $B(R) \subs \valus$ consist of all
valuations $k \in \valus$ such that, for all $s, t \in S$, if $s R t$
then $k(s) \leq k(t)$. 
We have the following result.

\begin{thm}{} \label{theo-logforkern}
Given a game structure $\game$,  the relation
$\priosim_1$ (resp.\ $\priobis_1$) can be characterized as the
largest (resp.\ largest symmetrical) relation $R$ such that, for all
states $s, t$ with $s R t$, we have $s \loceq t$ and 
\begin{align}
  & \forall k \in B(R) .
    \forall x_1 \in \dis_1(s) . 
    \exists y_1 \in \dis_1(t) .
    \forall y_2 \in \dis_2(t) .
  \exists x_2 \in \dis_2(s) .
    \bigl(
    \E_{t}^{y_{1},y_{2}}(k) \geq \E_{s}^{x_{1},x_{2}}(k)
    \bigr) \eqpun .  \label{eq-logforkern}
\end{align}
\end{thm}

\begin{proof}
The proof proceeds by induction on the computation of the fixpoint
relation $R$. 
We first present the case for $\priosim_1$.
Call $R_n$ the $n$-th iterate of the simulation relation $R$, 
and let $d_n$ be the $n$-th iterate of
$[\priosim_1]$, as in Theorem \ref{theo-picard}.
We prove by induction that, for all states $s, t \in S$, we have 
$s R_n t$ iff $d_n (s, t) = 0$. 
We define $d_0(s, t) = \metr{s \loceq t}$.
The base case is then immediate because $s R_0 t$ iff $d_0(s, t) = 0$.
Consider the induction step, for $n \geq 0$, 
and consider any states $s, t \in S$. 
Assume first that $d_{n+1}(s,t) > 0$: then, it is easy to show that we can
find a value for $k$ in (\ref{eq-logforkern}) that witnesses 
$(s, t) \not\in R_{n+1}$, since the constraints on $k$ due to $B(R_n)$
are weaker than those due to $C(d_n)$. 
Conversely, assume that there is a $k \in B(R_n)$ that witnesses
$(s,t) \not\in R_{n+1}$.
Then, by scaling all $k$ values so that they are all smaller than the
smallest non-zero value of $d_n(s',t')$ for any $s',t' \in S$, we can
find a $k' \in C(d_n)$ which also witnesses $d_{n+1}(s,t) > 0$, as
required.

The case for $\priobis_g$ is analogous, due to the similarity of the
Picard iterations (\ref{eq-picard-sim}) for $\priosim_1$ and
(\ref{eq-picard-bisim}) for $\priobis_g$.  
\end{proof}

We note that the above theorem allows the computation of $\priobis_g$
via a partition-refinement scheme. 
From the logical characterization theorem, we obtain the following
corollary. 

\begin{cor}{}
For any game structure $\game$ and states $s,t$ of $\game$, 
we have $s \priobis_g t$ iff $\sem{\varphi}(s) = \sem{\varphi}(t)$ holds for every $\varphi \in \qmu$ and
$s\priosim_1 t$ iff $\sem{\varphi}(s) \leq \sem{\varphi}(t)$ holds for every $\varphi \in \qmu^+_1$.
\end{cor}

%
%
%

\paragraph{Relation between Game Metrics and (Bi-)simulation Metrics.}

The a priori metrics assume an adversarial relationship between the players.
We show that, on turn-based games, the a priori bisimulation metric
coincides with the classical bisimulation metric where the players 
cooperate. 
%

We define such ``cooperative'' simulation and bisimulation metrics
$[\altsim_{12}]$ and $[\altbis_{12}]$ as the metric analog of
classical (bi)simulation \cite{Milner90,SL94}.
We define the metric transformers $H_{\altsim_{12}}: \metrsp \mapsto
\metrsp$ and $H_{\altbis_{12}}: \metrsp \mapsto \metrsp$, for all
metrics $d \in \metrsp$ and $s, t \in S$, by:
\begin{align*}
  H_{\altsim_{12}}(d)(s,t) & =
  \metr{s \loceq t} \imax
    \Sup_{k \in C(d)} \,
    \Sup_{x_1 \in \dis_1(s)} \,
    \Sup_{x_2 \in \dis_2(s)}  \,
    \Inf_{y_2 \in \dis_2(t)} \,
    \Inf_{y_1 \in \dis_1(t)} \,
    \{\E_s^{x_1,x_2}(k)\ -\ \E_t^{y_1,y_2}(k)\} \eqpun . \\[1ex]
  H_{\altbis_{12}}(d)(s,t) & = 
    H_{\altsim_{12}}(d)(s,t) \imax
    H_{\altsim_{12}}(d)(t,s) \eqpun .
\end{align*}
The metrics $[\altsim_{12}]$ and $[\altbis_{12}]$ are defined as
the least fixed points of $H_{\altsim_{12}}$ and $H_{\altbis_{12}}$ 
respectively. 
The kernel of these metrics define the classical probabilistic simulation and 
bisimulation relations.

\begin{thm}{}\label{theo-tb-conc-vs-12}
The following assertions hold. 
\begin{enumerate}[\em(1)]
\item 
On turn-based game structures, 
    $[\altbis_g] = [\altbis_{12}]$. 
\item 
There is a deterministic game structure $G$ and states $s,t$ in $G$
  such that 
  $\metr{s \altbis_g t} > \metr{s \altbis_{12} t}$. 
\item 
There is a deterministic game structure $G$ and states $s,t$ in $G$
  such that 
  $\metr{s \altbis_g t} < \metr{s \altbis_{12} t}$. 
\end{enumerate}
\end{thm}
\begin{figure}\centering
{\includegraphics[height=1in,width=3.5in]{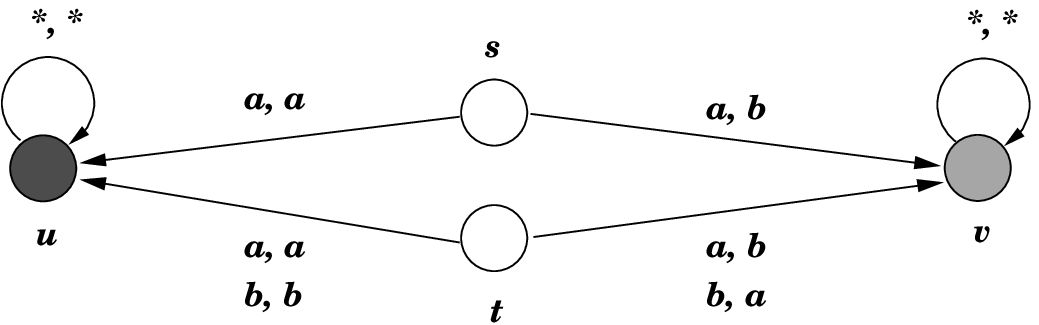}}
\caption{$\metr{s \altbis_g t} = \frac 1 2$ and 
$\metr{s \altbis_{12} t} = 0$
\label{fig:pl1vspl2alt}}
\end{figure}
\begin{figure}\centering
{\includegraphics[height=1in,width=3.5in]{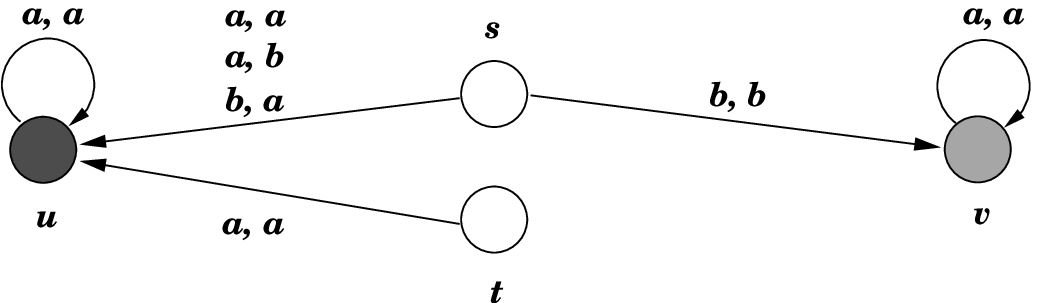}}
\caption{
  $\metr{s \altbis_g    t} = 0$
  but 
  $\metr{s \altbis_{12} t} =  1.$
  \label{figa-pl1vspl12sim}}
\end{figure}

\begin{proof}
For the first part, since we have turn-based games, only one player
has a choice of moves at each state.
We say that a state $s$ belongs to player~$\ii \in \set{1,2}$ if player
$\jj$ has only one move at $s$.
First, notice that due to the presence of the variable $\turn$, the
metric distance between states belonging to different players is
always~1, for all the metrics we consider.  
Thus, we focus on the metric distances between states belonging to the
same player. 
Consider two player~1 states $s, t \in S$. 
From the definitions of $H_{\priosim_1}$ and $H_{\priosim_{12}}$,
for $d \in \metrsp$, by dropping the moves of player~2, it is easy 
to see that $H_{\priosim_1}(d) = H_{\priosim_{12}}(d)$, and 
$H_{\priobis_g}(d) = H_{\priobis_{12}}(d)$.
Since this holds for any $d \in \metrsp$, it holds for the fixpoints,
$\metr{\altbis_g}$ and $\metr{\altbis_{12}}$.

The second part is proved by the game in Figure~\ref{fig:pl1vspl2alt},
where $\metr{s \loceq t} = 0$ and $\metr{u \loceq v} = 1$.
The latter yields $\metr{u \priobis_g v} = 1$.
Since player~1 has no choice of moves at state $s$, the maximum
probability with which player~1 can guarantee a transition to
either state $u$ or state $v$ is $0$.
But from state $t$, by playing moves $a, b$ with probability
$\frac 1 2$ each, player~1 can guarantee reaching states $u$
and $v$ with probability $\frac 1 2$, which implies that over
all $k \in C(d)$, given that $d(u, v) = 1$ from 
$\metr{u \priobis_g v} = 1$,
the maximum $k$ expectation that player~1 can guarantee is 
$\frac 1 2$.
Therefore $\metr{s \altbis_g t} = \frac 1 2$.
But if player~2 co-operates, then $\metr{s \altbis_{12} t} = 0$. 

The third part is proved by the
game in Figure~\ref{figa-pl1vspl12sim}
where again $\metr{s \loceq t}=0$ and $\metr{u \loceq v}=1$.
Since the players don't have any moves to transition to state
$v$ from state $t$, $\metr{s \altbis_{12} t} = 1$, whereas 
$\metr{s \altbis_g t} = 0$.
\end{proof}

If we consider Markov decision processes (MDPs), we have that on
$i$-MDPs, the metric $\altsim_i$ coincides with $\altsim_{12}$, since
player $\jj$ has no moves, for $\ii \in \set{1,2}$. 
On the other hand, the metric $\altsim_{\jj}$ provides no
information on $\altsim_{12}$.

\begin{thm}{}
The following assertions hold. 
\begin{enumerate}[\em(1)]
\item 
For $i$-MDPs 
    we have $[\altsim_i] = [\altsim_{12}]$.
\item 
There is a deterministic 2-MDP $G$ with states 
  $s,t$ such that 
  $\metr{s \altsim_1 t} < \metr{s \altsim_{12} t}$.
\item 
There is a deterministic 2-MDP  $G$ with states 
  $s,t$ such that 
  $\metr{s \altsim_1 t} > \metr{s \altsim_{12} t}$.
\end{enumerate}
\end{thm}
\begin{figure}[t]\centering
{\includegraphics[height=1in,width=3.5in]{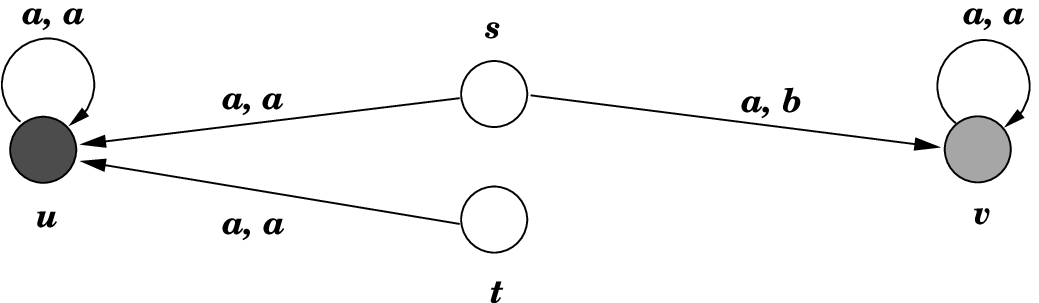}}
\caption{
  $\metr{s \altsim_1    t} = 0$ and 
  $\metr{s \altsim_{12} t} = 1.$
  Also, 
  $\metr{t \altsim_1    s} = 1$ and 
  $\metr{t \altsim_{12} s} = 0.$
  \label{figb-pl1vspl12sim}}
\end{figure}
\begin{proof}
From the definitions of $H_{\priosim_1}$ and $H_{\priosim_{12}}$,
restricted to MDPs, where only one player has a choice of moves,
the first assertion follows.

The second and third assertions are proved by the
deterministic 2-MDP in Figure~\ref{figb-pl1vspl12sim}, where again 
$\metr{s \loceq t}=0$ and $\metr{u \loceq v}=1$.
For the second assertion we note that since $d(u, v) = 1$,
for any choice of $k \in C(d)$, player~1 cannot get a higher 
expectation of $k$ from state $s$ when compared to state $t$,
because at state $s$, player~2 always has a move that will lead to 
a state yielding a lower $k$ expectation.
Therefore, $\metr{s \priosim_1 t} = 0$.
Further, for $k(v) = 1$ and $k(u) = 0$, which satisfies the
constraints on $k$, we have no moves for either player from state
$t$, which implies $\metr{s \priosim_{12} t} = 1$.

We prove the third assertion by showing that, for the 2-MDP of
Figure~\ref{figb-pl1vspl12sim}, we have 
$\metr{t \altsim_1 s} > \metr{t \altsim_{12} s}$
(which is the third assertion, with $s$ and $t$ exchanged). 
Note that when player~2 cooperates, 
the expectation of any $k \in C(d)$ from state $s$ is
always at least as much as the expectation from state $t$.
Thus $\metr{t \altsim_{12} s} = 0$.
Finally, there exists a $k \in C(d)$, with $k(u) = 1$ and 
$k(v) = 0$, for which $\metr{t \altsim_1 s} = 1$, which completes
the proof.
\end{proof}

\paragraph{Computation.}

We now show that the metrics are computable to any degree
of precision. 
This follows since the definition of the distance between two states
of a given game,
as the least fixpoint of the metric transformer (\ref{eq-priosim}), 
can be written as a formula in the theory of reals, which is decidable
\cite{Tarski51}. 
Since the distance between two states may not
be rational, we can only guarantee an approximate computation in general.

Without loss of generality, we assume that the states of $\game$ are 
labeled $\set{s_1,\ldots, s_n}$ for some $n\in\naturals$.
The construction is standard (see, e.g., \cite{dAM04}), we recapitulate the 
main steps.
We denote by ${\mathbf R}$ the real-closed field $(\reals, +, \cdot, 0, 
1, \leq)$ of the
reals with addition and multiplication.
An {\em atomic formula} is an expression of the form $p > 0$ or $p = 0$ where
$p$ is a (possibly) multi-variate polynomial with integer coefficients.
An {\em elementary formula} is constructed from atomic formulas by the grammar
\[
\varphi ::= a \mid
\neg \varphi \mid \varphi \wedge \varphi \mid \varphi \vee \varphi
\mid \exists x. \varphi \mid \forall x. \varphi,
\]
where $a$ is an atomic formula, $\wedge$ denotes conjunction, $\vee$ 
denotes disjunction,
$\neg$ denotes complementation, and $\exists$ and $\forall$ denote existential
and universal quantification respectively.
We write $\varphi \rightarrow \varphi'$ as shorthand for $\neg \varphi \vee \varphi'$.
The semantics of elementary formulas are given in a standard way 
\cite{ChangKeisler73}.
A variable $x$ is {\em free} in the formula $\varphi$ if it is not in the 
scope of a quantifier $\exists x$ or $\forall x$.
An {\em elementary sentence} is a formula with no free variables.
The theory of real-closed fields is decidable \cite{Tarski51}.

We introduce additional atomic formulas as syntactic sugar:
for polynomials $p_1$ and $p_2$, we write $p_1 = p_2$ for $p_1- p_2 = 0$, $p_1 > p_2$
for $p_1 - p_2 > 0$, and
$p_1 \geq p_2$ for $p_1 - p_2 = 0 \vee p_1 - p_2 > 0$.
Also, we write $p_1 \leq p_2$ for $p_2 \geq p_1$ and $p_1 < p_2$ for
$p_2 > p_1$.
Let $\vec{x},\vec{y}$ denote vectors of variables, where the dimensions of the vectors
will be clear from the context.
For $\sim \in \set{=,\leq,\geq}$,
we write $\vec{x} \sim \vec{y}$ for the
pointwise ordering, that is,
if $\bigwedge_i x_i \sim y_i$.
A subset $C \subseteq \reals^m$ is {\em definable} in ${\mathbf R}$ if there
exists an elementary formula $\varphi_C(\vec{x})$ such that for any $\vec{x}_0\in\reals^m$,
we have $\varphi_C(\vec{x}_0)$ holds in ${\mathbf R}$ iff $\vec{x}_0\in C$.
A function $f: \reals^k\rightarrow \reals^m$ is {\em definable} in ${\mathbf R}$ if there exists
an elementary formula $\varphi_f(\vec{y}, \vec{x})$ with free variables $\vec{y}$, $\vec{x}$ such
that for all constants $\vec{y}_0\in\reals^m$ and $\vec{x}_0 \in \reals^k$
the formula $\varphi_f(\vec{y}_0,\vec{x}_0)$ is true in $\mathbf{R}$ iff $\vec{y}_0 = f(\vec{x}_0)$.
We start with some simple observations about definability.

\begin{lem}{}\label{lem-definability}
(a) If functions $f_1: \reals^k \rightarrow \reals^m$ and $f_2:\reals^k\rightarrow \reals^m$ 
are definable in ${\mathbf R}$ then so are
the functions
\[ 
\begin{array}{rcl}
(f_1 - f_2)(\vec{x}) & = & f_1(\vec{x}) - f_2(\vec{x}) \\[1ex]
(f_1\sqcup f_2)(\vec{x}) & = & f_1(\vec{x}) \sqcup f_2(\vec{x})
\end{array}
\]
(b) If $f:\reals^{k+l}\rightarrow \reals^m$ is definable in ${\mathbf R}$, and $C\subseteq \reals^k$
is definable in ${\mathbf R}$,
then $(\Sup_C f) : \reals^l \rightarrow \reals^m$ defined as
\[
(\Sup_C f) (\vec{y}) = \Sup_{\vec{x}\in C} f(\vec{x},\vec{y})
\]
is definable in ${\mathbf R}$.
\end{lem}

\begin{proof}
For part (a), let $\varphi_1(\vec{y},\vec{x})$ and $\varphi_2(\vec{y},\vec{x})$ be formulas
defining $f_1$ and $f_2$ respectively.
Then, $f_1 - f_2$ is defined by the formula 
\[
\exists \vec{z}_1.\exists\vec{z}_2.
(\varphi_1(\vec{z}_1,\vec{x})\wedge \varphi_2(\vec{z}_2,\vec{x})
\wedge \vec{y} = \vec{z}_1 - \vec{z}_2),
\]
and
$f_1 \sqcup f_2$ is defined by the formula
\[
\exists \vec{z}_1.\exists\vec{z}_2.
(\varphi_1(\vec{z}_1,\vec{x})\wedge \varphi_2(\vec{z}_2,\vec{x})
\wedge \bigwedge_i \left[ 
(\vec{z}_{1,i} \geq \vec{z}_{2,i} \wedge \vec{y}_i = \vec{z}_{1,i}) \vee 
(\vec{z}_{1,i} < \vec{z}_{2,i} \wedge \vec{y}_i = \vec{z}_{2,i}) \right]) 
\eqpun .
\]
For part (b), let $\varphi_f(\vec{z}, \vec{x},\vec{y})$ define $f$, where $\vec{x}$ is of dimension $k$,
$\vec{y}$ of dimension $l$, and $\vec{z}$ of dimension $m$, respectively.
Let $\psi_C(\vec{x})$ define $C$.
Then, the following formula with free variables $\vec{z}$, $\vec{y}$
(call it $\varphi(\vec{z},\vec{y})$) states that $\vec{z}$ is an 
upper bound of $f(\vec{x},\vec{y})$ for all $\vec{x}\in C$:
\[
\forall \vec{x}_1. \forall \vec{z}_1. 
(\psi_C(\vec{x}_1) \wedge \varphi_f(\vec{z}_1,\vec{x}_1, \vec{y}) 
\rightarrow \vec{z}_1\leq \vec{z}),
\]
and $\sup_C f$ is defined by the formula with free variables $\vec{z}$, $\vec{y}$ given by:
\[
\varphi(\vec{z},\vec{y}) 
\wedge 
\forall \vec{z}_1. 
(\varphi(\vec{z}_1,\vec{y}) \rightarrow \vec{z}\leq \vec{z}_1) \eqpun .
\]
\end{proof}

\begin{thm}{}
\label{theo-compute-theory-of-reals}
Let $\game$ be a game structure and $s,t$ states of $\game$.
For all rationals $v$, and all $\epsilon > 0$, it is decidable
if $|[s \priosim_1 t] - v | < \epsilon$ and if $|[s \priobis_g t] - v | < \epsilon$.
It is decidable if $s \priosim_1 t$ and if $s \priobis_g t$.
\end{thm}

\begin{proof}
First, we use a result of Weyl \cite{Weyl50} that the minmax value of a
matrix game with payoffs in $\reals$
can be written as an elementary formula in the theory of real-closed fields.
This implies that for any state $s$, 
the function $\pre_1(\vec{k})(s)$ is definable in ${\mathbf R}$.
Also, for $d\in \metrsp$, the set $C(d)$
is definable in ${\mathbf R}$ (since conjunctions of linear constraints are
definable in ${\mathbf R}$).
Hence, by Lemma~\ref{lem-definability}(a) and~(b), we have
that $\sup_{\vec{k}\in C(d)} \bigl(\pre_1(\vec{k})(s) - \pre_1(\vec{k})(t)\bigr)$ is
definable for any metric $d\in \metrsp$, and states $s$ and $t$ of $\game$.
By another application of Lemma~\ref{lem-definability}(a), we have that
the function 
\[
  H_{\priosim_1}(d)(s,t) = 
  (s\equiv t) \sqcup \sup_{\vec{k}\in C(d)} \bigl(
  \pre_1(\vec{k}(s) - \pre_1(\vec{k})(t) \bigr) \eqpun .
\]
is definable for $d\in\metrsp$ and states $s$  and $t$ of $\game$.

Consider the set of free variables 
$\set{y(s,t), d(s,t) \mid s, t \in S}$, 
where $d$ is a vector of $n^2$ free variables defining the metric
$d$, and where $y$ is a vector of $n^2$ variables. 
Let $\varphi(y,d)$ be a formula in ${\mathbf R}$, with free
variables in the above set, such that
$\varphi(y,d)$ is true iff $y(s,t) = H_{\priosim_1}(d)(s,t)$ holds for
all $s,t \in S$.
Then the formula $\varphi^*(y)$ with free variables $y$, defined as:
\[
  \exists d. (\varphi(y,d) \wedge y = d),
\]
defines a fixpoint of $H_{\priosim_1}(d)$.
Finally, the formula $\psi(y)$, given by 
\[
\varphi^*(y) \wedge \forall y'. (\varphi^*(y') \rightarrow y\leq y') \eqpun .
\]
defines the least fixpoint of $H_{\priosim_1}$
(again, $y' = \set{y'(s,t) \mid s,t \in S}$ is a matrix of $n^2$
variables, and $y \leq y'$ iff $y(s,t) \leq y'(s,t)$ for all $s,t \in S$).
Thus, $\psi(y)$ is true iff $y(s,t) = [s\priosim_1 t]$ for all $s,t \in S$.

While this shows that $[s\priosim_1 t]$ is algebraic, there are
game structures $\game$ with all transition probabilities being rational, 
but with states $s$ and $t$ of $\game$ such that 
$[s\priosim_1 t]$ is irrational.
So, we use the formula above to approximate the value of $[s\priosim_1 t]$
to within a constant $\epsilon$. 
For states $s,t$ and rationals $v, \epsilon$, we have that
$|[s \priosim_1 t] - v| < \epsilon$ iff 
$\exists y.(\psi(y) \wedge |y(s,t) - v| < \epsilon)$ is valid, 
and this can be decided since ${\mathbf R}$ is decidable.

A similar construction shows that the question whether
$|[s \priobis_g t] - v| < \epsilon$, is decidable
for states $s,t$ and rationals $v, \epsilon$: we ensure that $y$ is 
a symmetric fixpoint
by conjoining to $\varphi^*(y)$ constraints $y(s,t) = y(t,s)$ for all states
$s,t$.

If the formula $\exists y. (\psi(y) \wedge y(s,t) = 0)$,
where we assert that the distance
between $s$ and $t$ is zero, is valid, we can conclude that
$s\priosim_1 t$.
This implies that the relation $s\priosim_1 t$ is decidable for
any game structure $\game$ and states $s$ and $t$ of $\game$. 
A similar construction for $\priobis_g$ shows that the relation
$s \priobis_g t$ is also decidable for any game structure $\game$ and
states $s,t$ of $\game$.
\end{proof}

\section{Discussion}
\label{sec-discussion}

Our derivation of $\priosim_i$ and $\priobis_g$, for $\ii \in
\set{1,2}$, as kernels of metrics, seems somewhat abstruse: most
equivalence or similarity relations have been defined, after all,
without resorting to metrics. 
We now point out how a generalization of the usual definitions
\cite{SL94,CONCUR98AHKV,DGJP99,DGJP02}, suggested in
\cite{luca-icalp-disc-03,rupak-thesis}, fails to produce the ``right''
relations.
Furthermore, the flawed relations obtained as a generalization of 
\cite{SL94,CONCUR98AHKV,DGJP99,DGJP02} are no simpler than our
definitions, based on kernel metrics. 
Thus, our study of game relations as kernels of metrics carries no
drawbacks in terms of leading to more complicated definitions. 
Indeed, we believe that the metric approach is the superior one for
the study of game relations. 

We outline the flawed generalization of 
\cite{SL94,CONCUR98AHKV,DGJP99,DGJP02} as proposed in 
\cite{luca-icalp-disc-03,rupak-thesis}, explaining why it would seem a
natural generalization. 
The alternating simulation of \cite{CONCUR98AHKV} is defined over
deterministic game structures. 
Player-$i$ alternating simulation, for $\ii \in \set{1,2}$, is the
largest relation $R$ satisfying the following conditions, for all
states $s, t \in S$: 
%
%
$s \rel t$ implies $s \loceq t$ and 
%
  $\forall x_\ii \in \mov_\ii(s) \qdot \exists y_\ii \in \mov_\ii(t) \qdot 
   \forall y_\jj \in \mov_\jj(t) \qdot \exists x_\jj \in \mov_\jj(s) \qdot 
   \ddest(s,x_1,x_2) \rel \ddest(t,y_1,y_2)$. 
%

The MDP relations of \cite{SL94}, later extended to metrics by
\cite{DGJP99,DGJP02}, rely on the fixpoint (\ref{eq-mdp-rel-fix}), 
where $\sup$ plays the role of $\forall$, $\inf$ plays the role of
$\exists$, and $R$ is replaced by distribution equality modulo $R$, or
$\sqsubseteq_R$. 
This strongly suggests --- incorrectly --- that equivalences for
general games (probabilistic, concurrent games) can
be obtained by taking the double quantifier alternation $\forall \exists
\forall \exists$ in the definition of alternating simulation, changing
all $\forall$ into $\sup$, all $\exists$ into $\inf$, and replacing
$R$ by $\sqsubseteq_R$. 
The definition that would result is as follows. 
We parametrize the new relations by a player $i\in\set{1,2}$, as
well as by whether mixed moves or only pure moves are allowed. 
For a relation $R \subs S \times S$, for $M \in \set{\mov, \dis}$,
for all $s, t \in S$ and $\ii \in \set{1, 2}$ consider the following
conditions:
\begin{enumerate}[$\bullet$]

\item {\bf (loc)} $s \rel t$ implies $s \loceq t$.

\item {\bf ($M$-$i$-altsim)} $s \rel t$ implies \\
  $\forall x_\ii \in M_\ii(s) \qdot \exists y_\ii \in M_\ii(t) \qdot 
   \forall y_\jj \in M_\jj(t) \qdot \exists x_\jj \in M_\jj(s) \qdot 
   \trans(s,x_1,x_2) \sqsubseteq_{R} \trans(t,y_1,y_2)$;

\end{enumerate}
We then define the following relations: 
\begin{enumerate}[$\bullet$] 

\item For $\ii \in \set{1,2}$ and $M \in \set{\mov, \dis}$, 
  {\em player-$i$ $M$-alternating simulation\/}
  $\postsim_i^M$ is the largest relation that satisfies (loc) and
  ($M$-$i$-altsim). 

\item For $\ii \in \set{1,2}$ and $M \in \set{ \mov, \dis}$, 
  {\em player-$i$ $M$-alternating
  bisimulation\/} $\postbis_i^M$ is the largest symmetrical relation that
  satisfies (loc) and ($M$-$i$-altsim). 
\end{enumerate}
Over deterministic game structures, the definitions of 
$\altsimdet_i$ and $\altbisdet_i$ coincide with the 
alternating simulation and bisimulation relations of \cite{CONCUR98AHKV}.
In fact, $\altsimdet_i$ and $\altbisdet_i$ capture the 
{\em deterministic\/} semantics of 
$\qmu$, and thus in some sense generalize the results of 
\cite{CONCUR98AHKV} to probabilistic game structures. 

\begin{thm}{} \label{theo-logical-relations} 
For any game structure $\game$ and states $s, t$ of $\game$, 
the following assertions hold: 
\begin{enumerate}[\em(1)]
\item 
$s \altbisdet_i t$ iff 
$\dsem{\varphi}(s) = \dsem{\varphi}(t)$ holds for every $\varphi \in \qmu_i$.
\item $s \altsimdet_i t$ iff
$\dsem{\varphi}(s) \leq \dsem{\varphi}(t)$ holds for every $\varphi \in \qmu_i^+$.
\end{enumerate}
\end{thm}

The following lemma states that $\altsimran_i$ and $\altbisran_i$ are
the kernels of $[\postsim_i]$ and $[\postbis_i]$, connecting thus the
result of combining the definitions of \cite{SL94} and
\cite{CONCUR98AHKV} with a posteriori metrics. 

\begin{lem}{} \label{lem-post-kernel}
For all game structures $\game$, all players $i \in \set{1,2}$, 
and all states $s, t$ of $\game$, we
have $s \altsimran_i t$ iff $[s \postsim_i t] = 0$, and 
$s \altbisran_i t$ iff $[s \postbis_i t] = 0$.
\end{lem}

We are now in a position to prove that neither the $\mov$-relations
not the $\dis$-relations are the ``canonical'' relations on general
concurrent games, since neither characterizes $\sem{\qmu}$. 
In particular, the $\dis$-relations are too fine, and the
$\mov$-relations are incomparable with the relations $\priosim_i$ and
$\priobis_g$, for $i \in \set{1,2}$. 
We prove these negative results first for the $\dis$-relations. 
They follow from Theorem~\ref{theo-metrics-diff}
and~\ref{theo-logical-charact-metrics}. 

\begin{thm}{}
The following assertions hold: 
\begin{enumerate}[\em(1)]

\item For all game structures $\game$, all states $s, t$ of $\game$,
  and all $i \in \set{1,2}$, we have that 
  $s \altsimran_i t$ implies $s \priosim_i t$, and 
  $s \altbisran_i t$ implies $s \priobis_i t$. 

\item There is a game structure $\game$, and states $s, t$ of $\game$,
  such that $s \priosim_i t$ but $s \not\altsimran_i t$. 

\item There is a game structure $\game$, and states $s, t$ of $\game$, 
  such that $\sem{\varphi}(s) = \sem{\varphi}(t)$ for all $\varphi \in \qmu$, 
  but $s \not\altbisran_i t$ for some $i \in \set{1,2}$. 
\end{enumerate}
\end{thm}

We now turn our attention to the $\mov$-relations, showing that they
are incomparable with $\priosim_i$ and $\priobis_g$, for $i \in
\set{1,2}$. 

\begin{thm}{} \label{thm:altvsgame}
The following assertions hold: 
\begin{enumerate}[\em(1)]

\item
There exists a deterministic game structure $\game$ and states $s,t$ 
of $\game$ such that $s \altsimdet_1 t$ but $s \not \priosim_1 t$, 
and $s \altbisdet_1 t$ but $s \not\priobis_g t$.

\item
There exists a turn-based game structure $\game$ and states $s,t$ of $\game$
such that $s \priosim_1 t$ but $s \not \altsimdet_1 t$.
and       $s \priobis_g t$ but $s \not \altbisdet_1 t$.

\end{enumerate}
\end{thm}

\begin{proof}
The first assertion is proved via the deterministic game in 
Figure~\ref{fig:altsimnogamesim},
where $\metr{s \loceq t}=0$ and $\metr{u \loceq v}=1$ and 
$\mov_1(s) = \mov_2(s) = \{a, b\}$ and $\mov_1(t) = \mov_2(t) =
\{a, b, c\}$. In the figure, we use the variables $x$ and $y$ to
represent moves: if player~1 and player~2 moves coincide, $u$ is the
successor state, otherwise it is $v$.
Thus, the game from $s$ is the usual ``penny-matching'' game; the game
from $t$ is a version of ``penny-matching'' with 3-sided pennies. 

\begin{figure}\centering
{\includegraphics[height=1.1in,width=3.5in]{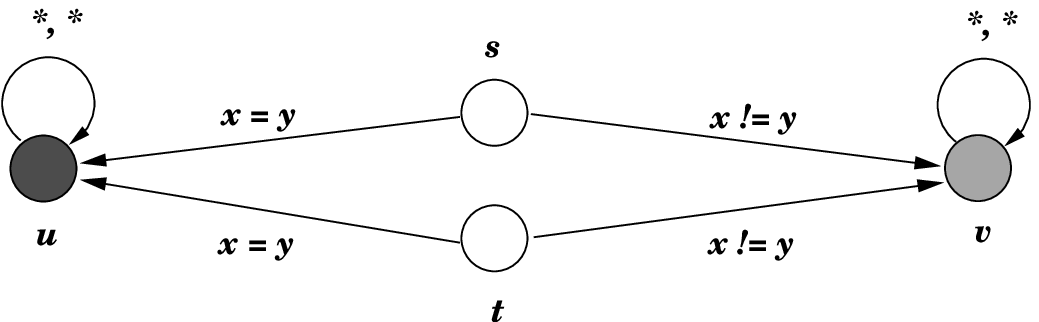}}
\caption{
$s \protect\sqsubseteq^\Gamma_1 t$ but $s \protect\not \protect\preceq_1 t$ and
         $s \protect\cong^\Gamma_1 t$ but $s \protect\not \protect\simeq_g t$}
\label{fig:altsimnogamesim}
\end{figure}

It can be seen that $s \altsimdet_1 t$.
On the other hand, we have $s \not \priosim_1 t$. 
Indeed, from state $s$, by playing both $a$ and $b$ with probability
$\frac{1}{2}$, player~1 can ensure that the probability of a
transition to $u$ is $\frac{1}{2}$.
On the other hand, from state $t$, player~1 can achieve at most
probability $\frac{1}{3}$ of reaching $u$ (this maximal probability is
achieved by playing all of $a$, $b$, $c$ with probability
$\frac{1}{3}$). 
The result then follows using
Theorem~\ref{theo-logical-charact-metrics}. 

The second assertion is proved via the game in 
Figure~\ref{fig:gamesimnoaltsim}. 
We have $s \not \altsimdet_1 t$: 
clearly, player-1's move $c$ at state $s$ cannot be mimicked
at $t$ when the game is restricted to pure moves.
On the other hand, we have $s \priosim_1 t$: 
since the move $c$ at $s$ can be imitated via the mixed move
that plays both $a$ and $b$ at $t$ with probability $\frac{1}{2}$
each, all $\qmu$ formulas have the same value, under $\sem{\cdot}$, at
$s$ and $t$, and the result follows once more using  
Theorem~\ref{theo-logical-charact-metrics}. 

\begin{figure}\centering
{\includegraphics[height=1.2in,width=3.5in]{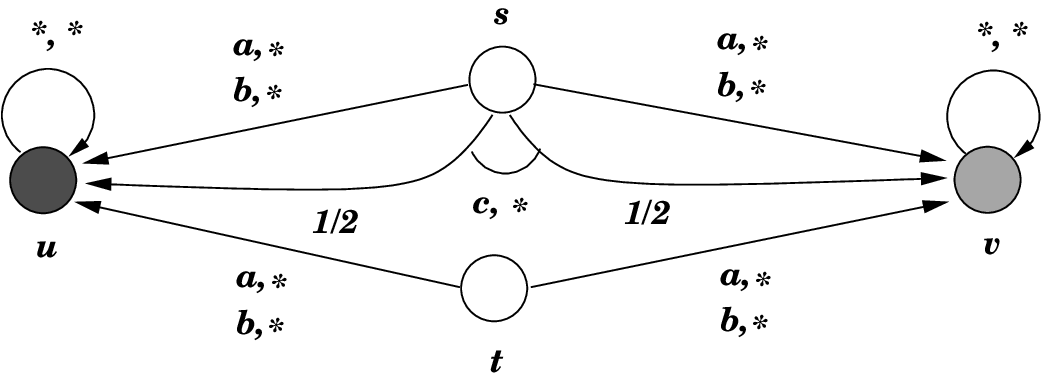}}
\caption{
$s \protect\preceq_1 t$ but $s \protect\not \protect\altsimdet_1 t$ and
         $s \protect\simeq_g t$ but $s \protect\not \protect\altbisdet_1 t$. }
\label{fig:gamesimnoaltsim}
\end{figure}
\end{proof}

Finally, we remark that, in view of
Theorem~\ref{theo-logforkern}, the definitions of the relations
$\priosim_i$ and $\priobis_g$ for $i \in \set{1,2}$ are no more
complex than the definitions of $\altsimran_1$, $\altsimdet_1$,
$\altbisran_1$, and $\altbisdet_1$.

\section{Conclusions}

We have introduced the metrics and relations that constitute the
natural generalizations of simulation and bisimulation to stochastic
games on graphs. 
These relations and metrics are tight, in the sense that the distance
between two states is equal to the maximum difference in value that
properties of the quantitative $\mu$-calculus can assume at the two
states: in other words, the relations characterize quantitative
$\mu$-calculus, in the same way in which ordinary bisimulation
characterizes $\mu$-calculus. 
The paper also provided a full picture of the connection between the
new metrics and relations, and the relations previously considered for
games. 

The main point left open by the paper concerns the algorithms for the
computation of the relations and metrics. 
The algorithms we provided rely on the decidability of the theory of
reals; it is an open question whether more efficient, and more direct,
algorithms exist, for the metrics or at least for the relations.

\subsection*{Acknowledgments.}
The first author was supported in part by the National Science
Foundation grants CNS-0720884 and CCR-0132780. 
The second author was supported in part by the National Science
Foundation grants CCF-0427202, CCF-0546170.
The fourth author was supported in part by the Netherlands Organization
for Scientific Research grant 642.000.505 and the EU grants IST-004527 and
FP7-ICT-2007-1 214755.

\bibliographystyle{plain}
\bibliography{main}

\end{document}